# Controlled doping of electrocatalysts through engineering impurities

*Se-Ho Kim*[1,†,*], *Su-Hyun Yoo*[1,†], *Sangyong Shin*[2], *Ayman A. El-Zoka*[1], *Olga Kasian*[1,3], *Joohyun Lim*[1,4], *Jiwon Jeong*[1], *Christina Scheu*[1], *Jörg Neugebauer*[1], *Hyunjoo Lee*[2], *Mira Todorova*[1], *Baptiste Gault*[1,5*]

[1]Max-Planck-Institut für Eisenforschung GmbH, Max-Planck-Straße 1, Düsseldorf 40237, Germany

[2]Department of Chemical and Biomolecular Engineering, Korea Advanced Institute of Science and Technology, Daejeon 34141, South Korea

[3]Helmholtz-Zentrum Berlin GmbH, Helmholtz Institut Erlangen-Nürnberg, Berlin 14109, Germany

[4]Department of Chemistry, Kangwon National University, Chuncheon 24342, Republic of Korea

[5]Department of Materials, Royal School of Mines, Imperial College, London, SW7 2AZ, United Kingdom

[†] these authors contributed equally.

[*] co-corresponding authors.

**Fuel cells recombine water from $H_2$ and $O_2$ thereby powering *e.g.* cars or houses[1] with no direct carbon emission. In anion-exchange membrane fuel cells (AEMFCs), to reach high power densities, operating at high *p*H is an alternative to using large volumes of noble metals catalysts at the cathode, where the oxygen-reduction reaction occurs[2]. However, the sluggish kinetics of the hydrogen-oxidation reaction (HOR) hinders upscaling despite promising catalysts[3–5]. Here, we observe an unexpected ingress of B into Pd nano-catalysts synthesised by wet-chemistry, gain control over this B-doping, and report on its influence on the HOR activity in alkaline conditions. We rationalize our findings using *ab-initio* calculations of both H- and OH-adsorption on B-doped Pd. Using this 'impurity engineering' approach, we thus design Pt-free catalysts as required in electrochemical energy conversion devices, *e.g.* next generations of AEMFCs, that satisfy the economic and environmental constraints, *i.e.* reasonable operating costs and long-term stability, to enable the 'hydrogen economy'.**

*Green hydrogen* is produced by electrolyzers powered by solar[6,7], wind[8], geothermal[9], or tidal[10] renewable energy. Whenever needed, $H_2$ is supplied to fuel cells to generate electricity with an efficiency reaching 93.5%[11] and no direct carbon emissions. However, to compete with fossil fuel-based power generation, these electrochemical energy conversion devices need new materials that are affordable and durable. In alkaline fuel cells[12], non-noble metal-based electro-catalysts for the cathodic oxygen reduction reaction with activity and durability comparable or superior to the scarce Pt have been reported[13,14]. However, the kinetics of the anodic HOR in alkaline conditions is too slow, which is true also for Pt-group metal catalysts[15]. Pd nano-catalysts with oxophilic $CeO_x$ exhibited the highest recorded HOR specific exchange current (51.5 mA $mg^{-1}_{Pd}$)[5] in alkaline electrolytes. The high loads of precious metals required hinder upscaling the commercial development of efficient AEMFCs[16,17] and it motivates the search for new materials.

We recently reported that using sodium borohydride ($NaBH_4$) as reducing agent in room temperature synthesis of free-standing Pd nano-catalysts led to an ingress of impurities from the aqueous solution, *i.e.* Na, K[18]. The absence of surfactant, used to avoid agglomeration, leads to the formation of a complex aggregated structure previously referred to as a metallic nano-aerogel (MNA)[19,20]. Ever since its discovery during World War II, $NaBH_4$ has been widely used in synthetic chemistry[21]. Its excellent reducing properties result in lower operation input, *e.g.* no heating or additional organics are required to reduce metallic precursors. $NaBH_4$ has commonly been used to synthesize metal nanoparticles for catalytic[22], antimicrobial[23], electrochemical[24], and optical[25] applications. However, the concentration of B impurities in the products is rarely considered[26], and not systematically studied, yet the presence of B increases the lattice parameter[27] and could greatly modify the material's properties.

Here, we predict the stable integration of B within growing Pd-crystals and use theoretical guidelines to design a set of Pd-MNAs, with adjustable composition. We assess their performance towards the HOR in alkaline conditions and demonstrate the effect of the ingress of B from the solution into the material. Our impurity-doping approach could be generalised to help design *in-silico* future catalysts with enhanced activity by estimating the adsorption properties of various elements and their influence on the H- and OH-binding energies.

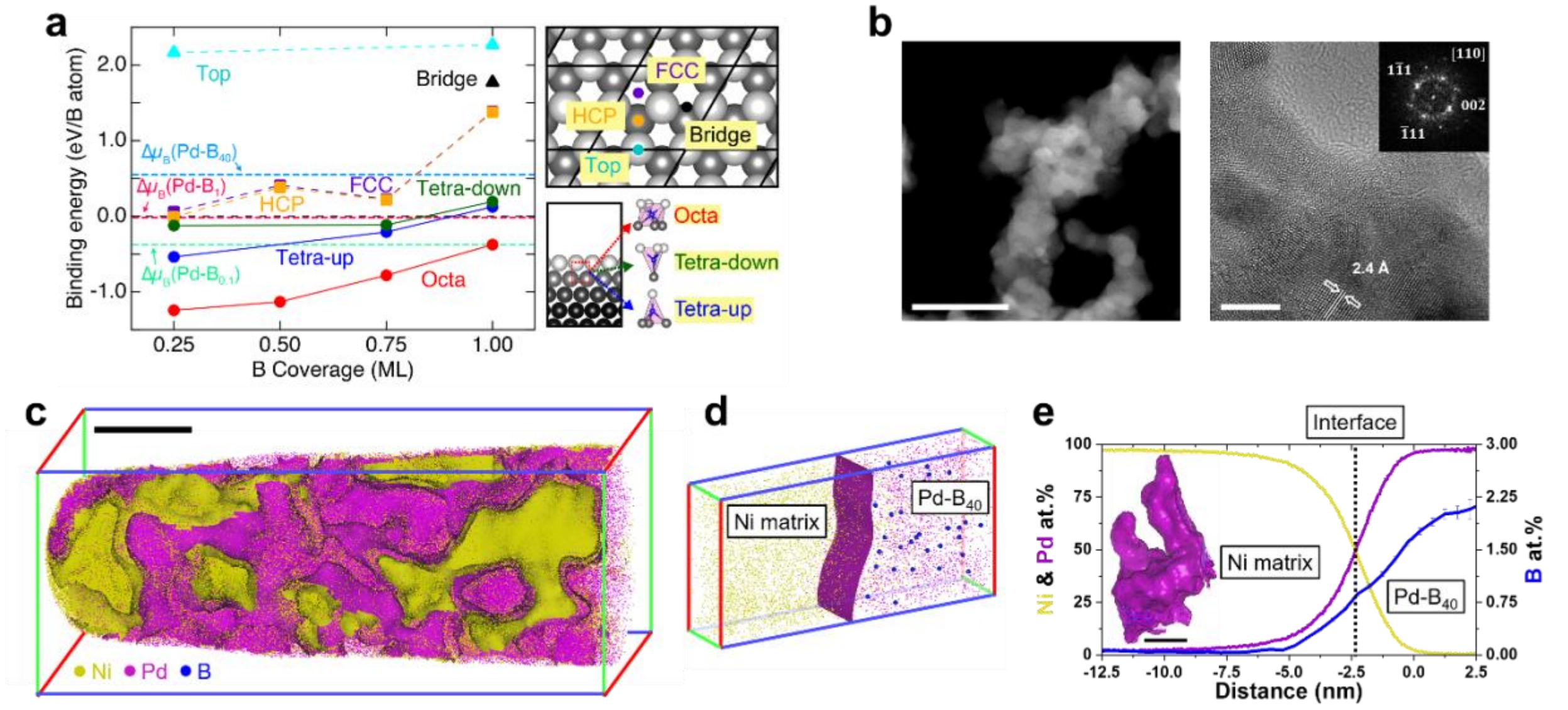

**Fig. 1. DFT calculation and high-resolution microscopy and microanalysis of B-incorporated Pd nano-materials.** (a) Binding energies $E_b$ of B adsorbates at the Pd(111) surface and sub-surface for several adsorbate coverages in the range from 0.25 ML to 1 ML. Each coloured solid line corresponds to a different B binding site. Different chemical potentials of $B_x$ (x gives the reductant to precursor ratio) of the considered synthesis conditions are shown as horizontal coloured dashed lines. The possible binding sites at the surface are shown (top) in top-view and at the sub-surface (bottom) in a side-view. White (grey) balls represent surface (sub-surface) Pd atoms. The $p(2\times2)$ surface unit cells are indicated by black solid lines. (b) HAADF-STEM (left) and HR-TEM (right) images of as-synthesized Pd-$B_{40}$ nanoparticles. Scale bars are 50 nm and 5 nm, respectively. The inset shows the [110] zone axis in a FFT pattern. (c) 3D atom maps of Pd-$B_{40}$ fully embedded in a Ni matrix as indicated by the Ni iso-surfaces. The scale bar is 20 nm. (d) Extracted region ($2\times5\times10$ nm$^3$) of interest along the Pd/Ni interface. Yellow, purple, and blue

dots mark the reconstructed positions of Ni, Pd, and B atoms, respectively. (e) 1D compositional profiles of elements across Pd-$B_{40}$ nanoparticles. Inset image shows a 3D reconstruction of the Pd-$B_{40}$. The scale bar is 20 nm.

First, we performed density-functional theory (DFT) calculations (see Methods for computational details) to calculate the B binding energies on a Pd(111) surface. Fig. 1a reports the respective energies for different high symmetry adsorption sites at and below the surface for increasing B coverage up to 1 monolayer (ML), where the number of adsorbates equals the number of substrate atoms in the first layer. Over this entire coverage range, the interstitial octahedral sub-surface site ($B_{Octa}$) is the most favourable, and B should be predominantly in the sub-surface rather than the surface region of Pd.

To prove this hypothesis, we mixed 0.4 M of $NaBH_4$ and 0.01 M of $K_2PdCl_4$, *i.e.* a ratio of reductant to precursor of (R:P) 40, to synthesize a Pd-MNA referred to as Pd-$B_{40}$ based on the nomenclature introduced in Ref.[18]. High-angle annular dark-field scanning- and high-resolution transmission electron microscopy (HAADF-STEM and HR-TEM) in Fig. 1b show an average ligament thickness of approx. 15 nm. The fast Fourier transform (FFT) pattern given as inset proves the face-centred cubic (FCC) structure of Pd. No particular elemental signal except Pd is obtained for the Pd-$B_{40}$ by X-ray photoelectron spectroscopy (XPS, see Fig. S1-S3). We then use atom probe tomography (APT) to evaluate the composition and elemental distribution in Pd-$B_{40}$ (see Methods and the Supplementary Information (SI)). Fig. 1c shows the reconstructed 3D atom map. A set of iso-surfaces delineates regions containing at least 50 at.% Pd (purple) and 75 at.% Ni (yellow), highlighting the gel and the Ni-matrix that embed the nano-catalyst to facilitate specimen preparation[28,29]. The ligaments size is comparable to HAADF-STEM observations. A slice through the APT point cloud, Fig. 1d, shows that B (in blue) is located inside Pd-$B_{40}$, not segregated to its surface, and the 1D composition profile, Fig. 1e, confirms the ingress of over 2 at% B within the Pd nano-catalyst. A nearest-neighbour analysis indicates no B clustering tendency within the resolution limits of APT[30,31], Fig. S13.

Incorporating chemical potentials $\Delta\mu_B$ into the modelling (*cf.* Fig. 1a) allows us to determine which adsorption sites are thermodynamically accessible. For $\Delta\mu_B$(Pd-$B_{40}$), corresponding to R:P 40 experimental growth conditions, the on-surface adsorption sites become thermodynamically metastable. These sites are at least 1 eV/(B atom) less favourable than sub-surface adsorption. If penetration into the sub-surface region is kinetically hindered they become the relevant adsorption sites. Using the value at the intersect of the chemical potential and binding energy lines (Fig. 1a) we calculate the achievable equilibrium sub-surface B concentration as roughly 1.85 B per Pd atom, which is substantially higher than the experimentally determined value. This, together with the observation that the experimental and theoretical concentration of B on the surface roughly agree, suggests that B incorporation into the nanoparticle commences by B atoms binding to the surface being continuously overgrown by subsequent Pd layers, resulting in a homogeneous distribution (Fig. S14).

To further confirm this hypothesis, we immersed a pure Pd wire (99.99+%) into 1.0 M $NaBH_4$ solution (~100 mole ratio) for 1h. The following APT analysis of its surface detected only trace amounts of B (<0.001 at.%) in the Pd (details in Fig. S15-S16) indicating that B is indeed integrated during the nanoparticle nucleation/growth process[32]. The B distribution inside the Pd catalyst can hence be controlled by tuning the synthesis conditions, which enables us to control the B-doping level by exploiting the B ingress to our advantage.

The 0.4 M $NaBH_4$ solution results in substantial B-doping of the Pd MNA. Hence, to confirm our expectations, we synthesise two new batches: Pd-$B_1$ (1 mole ratio) and Pd-$B_{0.1}$ (0.1 mole ratio).

Fig. 2a and 2b show 3D APT atom maps and composition profiles from the analysis of Pd-$B_1$ and Pd-$B_{0.1}$, respectively. The morphology and size are in both cases similar to Pd-$B_{40}$ (Fig. S17-S19). Using X-ray diffraction (XRD), we confirm that all samples are a single phase with a FCC structure (Fig. S21-S22). Neither boride nor oxide-related phases were found, but the insertion of interstitial B leads to lattice parameter expansion[27], resulting in an B-concentration related increase of the lattice parameters of B-doped Pd nanoparticles (Fig. 2c) compared to the pure-Pd lattice parameter reported to be 0.3890 nm[33] at room temperature.

A lower molar ratio of B during synthesis corresponds to a lower chemical potential of B in solution (*cf.* Fig. 1a), which results in a weakening and destabilisation of B binding in on-surface Pd sites with the binding energy for the HCP on-surface site: $E_b$ = -0.57, 0.00, and 0.36 eV/(B atom) for Pd-$B_{40}$, Pd-$B_1$, and Pd-$B_{0.1}$ at 0.25 ML, respectively). In agreement with our expectations of an effectively decreasing B-doping level in the Pd MNA, we find a B content of 2.7, 1.3, and 0.47 at.% in Pd-$B_{40}$, Pd-$B_1$, and Pd-$B_{0.1}$, respectively. Note that mole ratio above 1 is commonly reported for synthesizing nanoparticles[34,35], so regardless of the $NaBH_4$ concentration used during synthesis, most nanoparticles must contain B.

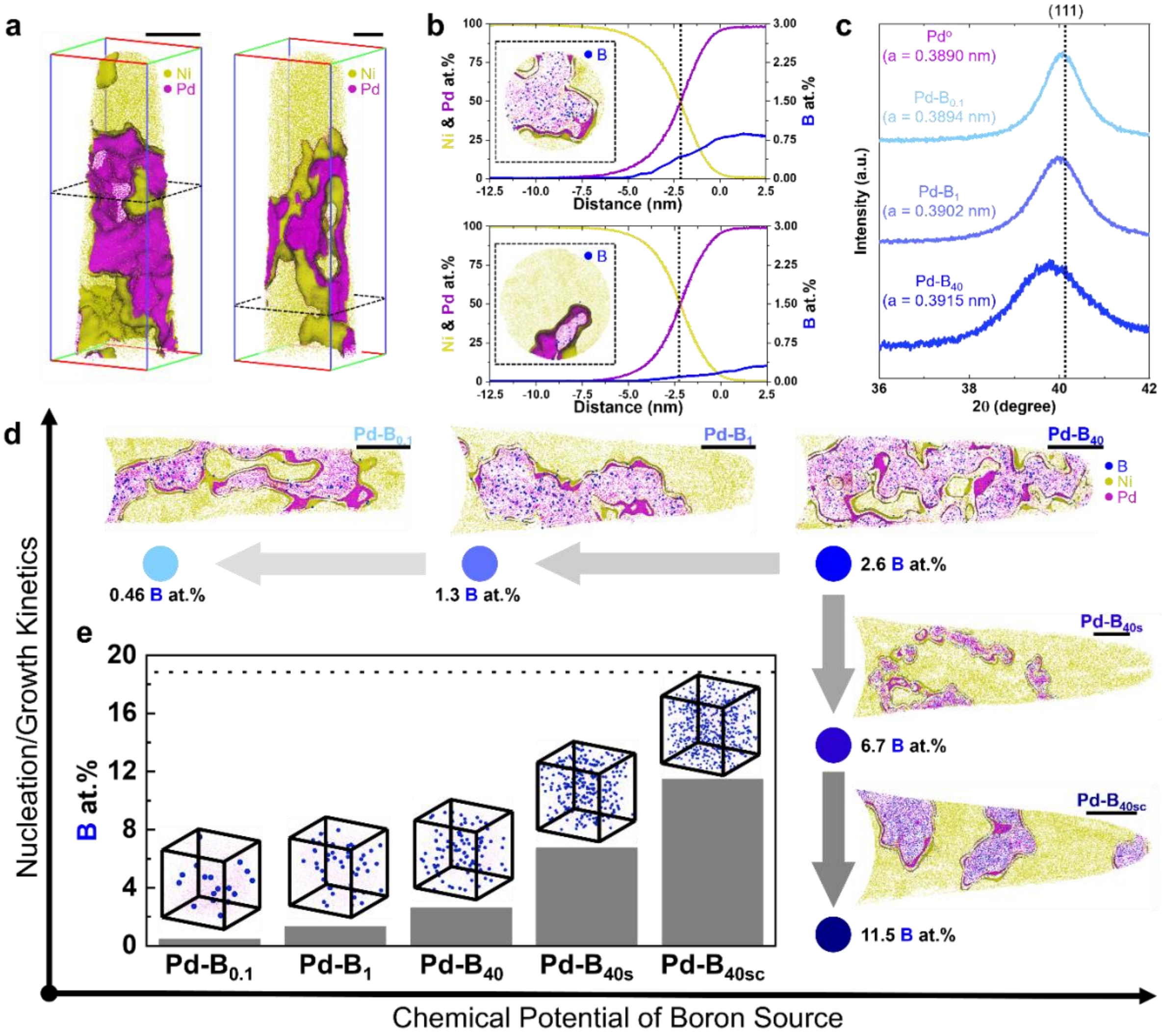

**Fig. 2. Controlling the magnitude of B content in Pd nanoparticles by synthesis.** (a) 3D atom maps of (left) Pd-$B_1$ and (right) Pd-$B_{40}$ nanoparticles embedded in a Ni matrix. (b) 1D compositional profile along the Pd/Ni interface for (top) Pd-$B_1$ and (bottom) Pd-$B_{40}$. Each inset shows the corresponding top-view tomogram. (c) XRD patterns of Pd-$B_x$ samples synthesized with different reductant concentrations. The dotted line represents the (111) peak position. (d) Schematic plot of the chemical potential of the boron source *vs.* Pd-$B_x$ nanoparticle growth kinetics. Insets show tomograms (one-nanometer thin slice along the x-axis) of each reconstructed Pd nanoparticle sample. All scale bars are 20 nm. (e) B atomic content in Pd-$B_{0.1}$, Pd-$B_1$, Pd-$B_{40}$, Pd-$B_{40s}$, and Pd-$B_{40sc}$ samples. The dotted line represents the maximum B content that FCC Pd metal can absorb without undergoing a phase

transformation[36]. Insets show the extracted cuboidal region of interest (5×5×5 $nm^3$) of each as-synthesized Pd-$B_x$ nanoparticle.

Besides varying the chemical potential, we can also adjust the reaction conditions, since B is progressively integrated during synthesis. Several levers could be used, for example, using *N*,*N* dimethylformamide (DMF), an organic solvent, and carbon black as a support in an ice-cooled solution, Li et al.[37] achieved 20 at.% doping of B in Pd nanoparticles. Here, we simply injected the reducing agent manually drop-by-drop at a relatively slow speed into the solution (approx. 50 µL $sec^{-1}$), allowing more time for B-integration, and achieved 6.7 at.% B (Pd-$B_{40s}$). Upon subsequent cooling of the solution to a temperature of 5 $^{o}$C, we obtained 11.5 at. % B (Pd-$B_{40sc}$). In each case, we used the same concentration of reducing agent as for the Pd-$B_{40}$ sample, however, the slower growth rate resulted in 2.5 – 4.3 times more B being integrated into the Pd-MNA (see details in SI). The morphology and size are similar to the other Pd-B samples, and neither B clustering nor surface-segregation is observed (Fig. S23-S24). Neither batch reached a B saturation in Pd of 18.6 at.% as in bulk-Pd at room temperature[36] or 12.7 at.% as suggested by the DFT calculations (Fig. S20). The above demonstrates that by precisely controlling the kinetics and chemical potential of B, we can control the B doping level, as schematically summarised in Fig. 2d.

Why doping Pd matters? Doping can be exploited to tailor the kinetics, selectivity, and stability of catalyst towards specific reactions[38,39]. As a platinum-group metal, Pd has similar electronic properties to Pt, but its stronger H absorption and adsorption behaviour result in slow reaction rates in acidic-HOR[15]. Lattice distortions from the presence of B in bulk Pd[40] change its physical properties and B-doping modifies its catalytic activity[41,42]. Specifically for the alkaline-HOR, the activity originates from a delicate balance of the adsorption strengths of H and OH at the catalyst's surface[43–45].

To estimate the strength of the chemical bond between the H adsorbate and the (B-doped) metal surface, we use the *d*-band model[46]. This model assumes that the strength of the bond is determined by the filling of the antibonding states: the higher the *d*-band centre energy (*i.e.* the closer it is to the Fermi energy), the stronger the chemical bond between adsorbate and substrate. Our analysis of the density-of-states (DOS) and electron density of the considered surfaces (Fig. S25), reveals a downward shift of the *d*-band centre. This is due to an accumulation of electron charge in the vicinity of the sub-surface B atoms and results in weaker H-binding on the Pd(111) surface with increasing B coverage.

Hydrogen binding energies, $E_b$(H), calculated for the B-free Pd(111) surface agree with other theoretical[47–50] and experimental[51] reports (see SI). The H binding energies calculated for varying coverages of H in FCC co-adsorbed with sub-surface B, are plotted in Fig. 3a. For 0, 0.25, and 1 ML $B_{Octa}$ and 1 ML H coverage we obtain -0.52, -0.37, and 0.32 eV/(H atom), respectively. For reference, we show in the Fig. also $E_b$(H) for 0.25 ML H in the FCC site on the pristine Pt(111) , *i.e.* -0.46 eV/(H atom)[52]. Also shown are the calculated OH binding energies $E_b$(OH) for pristine and for B-doped Pd(111). For these we observe a downward shift of the *d*-band centre position for increasing B sub-surface doping (*cf.* Table S10-S11), which strengthens the surface binding energy of OH in contrast to that of H [*e.g.*, $E_b$(0.5 ML OH at HCP sites) = -2.17, -2.22, and -2.30 eV for 0, 0.25, and 1 ML $B_{Octa}$ co-adsorption, respectively]. These results imply that there is an optimal level of doping for which the catalytic activity is boosted due to a weaker interaction of H with the surface at a point where the OH is also not strong.

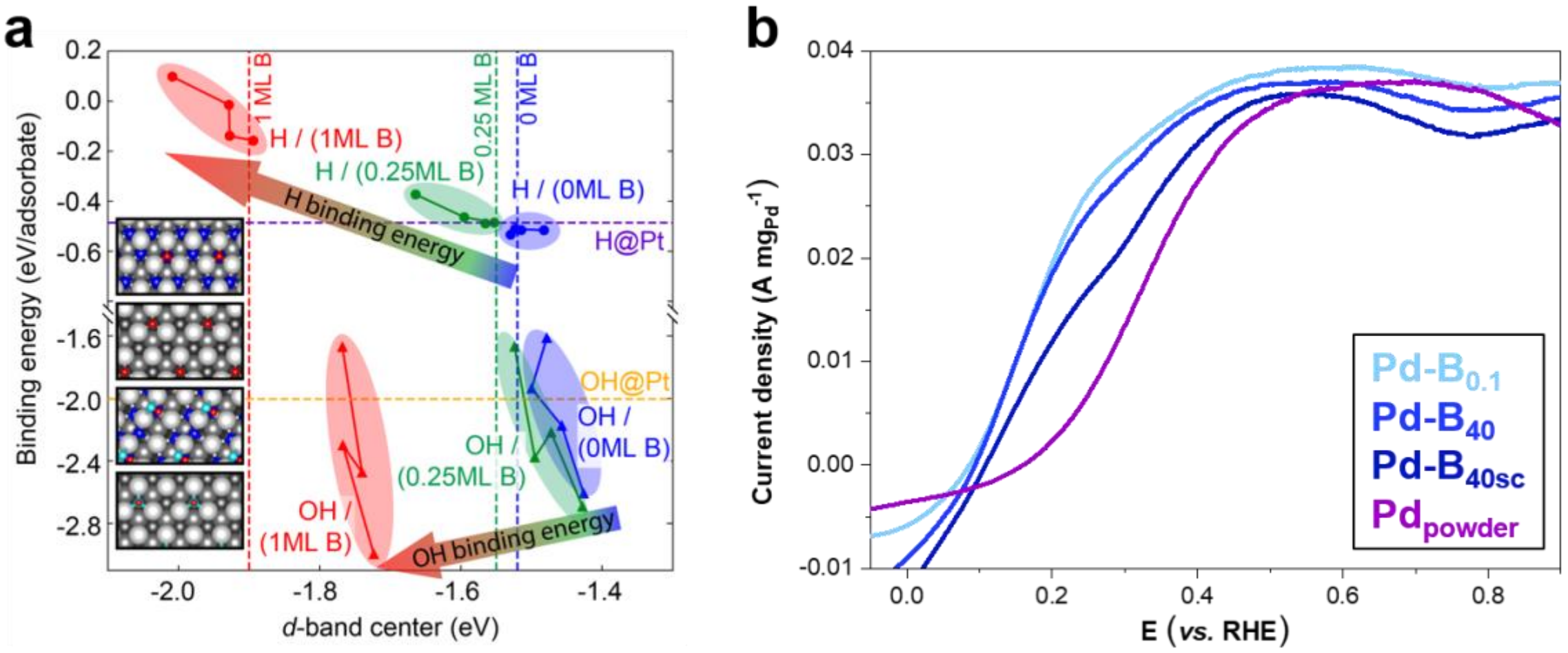

**Fig. 3. HOR binding energy calculations and experiments.** (a) Calculated H (top part) and OH (bottom part) binding energies plotted as a function of the calculated *d*-band centres referenced to the Fermi energy. Blue, green, and red lines indicate binding energies for 0.25 to 1 ML H (circle) and OH (triangle) in the presence of 0 ML, 0.25 ML and 1 ML of $B_{octa}$ in the sub-surface of Pd(111), respectively. Horizontal purple and orange dashed lines show the H and OH binding energies for a coverage of 0.25 ML at the most favourable Pt(111) hollow site at the taken from Ref. [52] and Ref. [53]. Vertical dashed lines indicate the calculated *d*-band centres of only B-doped Pd(111) surfaces. Insets depict H (red) and OH (O: cyan) adsorption on Pd (grey) structures with and without B (blue). (b) HOR LSV curve of the as-synthesized Pd-$B_{0.1}$, Pd-$B_{40}$, and Pd-$B_{40sc}$ samples and a reference $Pd_{powder}$. HOR was performed in a $H_2$-saturated 0.1M KOH electrolyte with 1600 rpm rotating speed. The LSV curve was measured from 1 V to -0.1 V (*vs* RHE) at a scanning rate of 10 mV $sec^{-1}$.

To verify our hypothesis, we measured linear sweep voltammetry (LSV) of HOR for Pd-$B_{0.1,40,40sc}$ and $Pd_{powder}$ in $H_2$-saturated 0.1 KOH solution (Fig. 3b). Measured current was normalized with the Pd mass of catalysts. We confirmed that a low level of B-doping improves the activity compared to $Pd_{powder}$, as expected from the decrease in the H surface binding energy, yet at higher doping levels, the activity decreases, which can be attributed to the increased surface binding energy of OH.

To summarise, we confirmed a predicted but undesired and uncontrolled incorporation of B arising from the solution during synthesis and proposed, and experimentally confirmed, an approach to design doped metallic nano-catalysts, with a clear influence on their catalytic activity towards the alkaline-HOR. We demonstrated how to engineer concentration of these impurities and control the doping level in the nanoparticles by adjusting the chemical potential and kinetics during synthesis. The presence of sub-surface B lowers the binding energy of H on the Pd(111) surface and leads to an increased reaction kinetics at low doping levels, despite the relatively strong increase in the surface binding energy of OH. B-doping hence enables a clear enhancement of the HOR kinetics yet the subtle balance between H and OH binding energies is necessary to design a superior HOR catalyst. Our approach provides a straightforward route to design catalysts that are efficient and stable for the green hydrogen-cycle. Its key idea, tuning impurity ingress from the synthesis solution that can be predicted from *ab-initio* DFT calculations, can be expanded to select impurities based upon their influence on the H- and OH-surface binding energies, but also on their sub-surface integration in order to ensure the doping of the catalyst.

# Methods

**Synthesizing Pd nanoparticles.** 0.01 M of Pd ions solution was prepared by dissolving potassium tetrachloropalladate (99.99%, Sigma-Aldrich) in 5 mL of distilled water. A white sodium borohydride (99.99%, Sigma-Aldrich) powder was dissolved in 5 mL of distilled water according to different mole ratios ($[NaBH_4]/[K_2PdCl_4]$ = 0.1 to 40). Then, two solutions were immediately mixed. After bubbles completely stopped, Pd nanoparticles were collected by centrifuging at 5000 rpm for 15 min. Pd nanoparticles were then re-dispersed in distilled water followed by centrifuging process. Washing process was done for thrice to remove any residuals. The collected Pd nanoparticles were then dried in a vacuum desiccator for 24 hr.

**XRD Analysis.** Powder X-ray diffraction was performed with RIKAKU SmartLab 9kW. The diffraction pattern of as-synthesized Pd samples were measured from the dried nanoparticles powders in 2θ mode using Co Kα radiation (λ = 1.54059 Å) and sampling step of $0.01^{o}$ at scan speed $2^{o}$ $min^{-1}$.

**XPS analysis.** XPS measurements were performed by applying a monochromatic Al Kα X-ray source (1486.6 eV) operating at 15 kV and 25 W. C 1s signal at 285.0 eV was used for a reference binding energy scale. The acquired spectra data was then analysed with the Casa XPS (http://www.casaxps.com/) software.

**TEM characterization.** HR-TEM and HAADF-STEM were performed using two different ThermoFisher Titan Themis 60-300 instruments operated at 300 kV, one with an image-corrector and one with a probe-corrector. The chemical composition of each $Pd\text{-}B_x$ sample was analysed by electron energy loss spectroscopy (EELS) and energy-dispersive X-ray spectroscopy (EDS) in the STEM mode. For Pd-$B_{40s}$ and Pd-$B_{40sc}$ samples, TEM and HAADF-STEM analyses were performed inside a JEM-2200FS TEM (JEOL) operating at 200 kV.

**APT characterization.** The Pd nanoparticle sample was embedded into a Ni matrix using a electrodeposition process as shown in Fig. S7. Xe-plasma focused-ion-beam (pFIB) (ThermoFischer Helios, Eindhoven) was performed to fabricate a needle-shape APT specimen from the co-deposited sample following a standard APT specimen preparation[54]. The final APT specimen is shown in Fig. S7c. Needle-shaped specimens were then loaded inside a Cameca LEAP 5000 XS system. APT measurements were performed in pulsed UV laser mode at a detection rate of 1 %, a laser pulse energy of 60 pJ, and a pulse frequency of 125 kHz. The specimen temperature was set to 50 K throughout the analysis. Data reconstruction and analyses were performed using the commercial software Imago visualization and analysis system standard (IVAS) 3.8.4 developed by Cameca Instruments. All 3D atom maps presented in this paper were reconstructed using the standard voltage reconstruction protocol[55].

**Electrochemical Measurements for HOR in alkaline.** Electrochemical measurements were performed with a CHI 760e potentiostat at room temperature. Non-doped Pd particles (99.9%, Sigma-Aldrich; < 1μm in size) were used as a reference labelled as $Pd_{powder}$. The catalyst ink was prepared by dispersing 5 mg of the catalyst in 2 ml of isopropyl alcohol and 6.66 μl of Nafion (5 wt.% sol. Sigma Aldrich) for 20 min sonication. Subsequently, 7.5 μl of the ink was drop-casted on the glassy carbon disk electrode (surface area = 0.247 $cm^2$). The amount of drop-casted catalyst was appropriate to completely cover the disk electrode. A double junction Hg/HgO filled with 1M KOH was used as a reference electrode and a carbon rod was used as a counter electrode. The as-prepared electrode was inserted into a 100 ml solution of 0.1 M KOH in which Ar gas was purged for 30 min. Cyclic voltammetry (CV) was performed at 50 mV $sec^{-1}$ scan rate (0.4 V to 1 V (*vs*

RHE)) three times, then 0.6 V (*vs* RHE) was applied to the working electrode and $H_2$ gas was bubbled for 30 min. The HOR LSV curve was measured from 1 V to -0.1V (*vs* RHE) at a scanning rate of 10 mV $sec^{-1}$. The experiment was repeated three times and the results were averaged.

**DFT calculation.** All presented DFT calculations are performed using the Vienna *Ab initio* Simulations Package (VASP)[56,57] with the projector augmented wave (PAW) approach.[58] The kinetic-energy cutoff employed for the plane-wave basis set is 500 eV. A Γ-centred (8×8×8) *k*-point grid is used for Brillouin-zone integrations for FCC Pd bulk and a (8×8×1) grid is used for the Pd(111) *p*(1×1) surface unit cell as the Pd(111) is the most favourable surface plane for FCC Pd[59]. Equivalently folded *k*-point meshes are used for larger surface cells. For the exchange-correlation approximation, the generalized gradient approximation (GGA) due to Perdew, Burke, and Ernzerhof[60,61] is used. A total of 16 (for Pd) and 3 (for B) electrons are treated as valence, respectively. Electronic and ionic relaxations are carried out until the total energy convergence is less than $10^{-5}$ eV per system, respectively $10^{-4}$ eV per system. With this setup, we obtain a lattice parameter $a$ = 3.959 Å and a cohesive energy $E_{coh}$ = 3.63 eV for Pd FCC bulk, in good agreement with theoretical[59,62] and experimental[63] results.

For the surface models, a symmetric supercell slab approach is used with an 18 Å vacuum region. Pd(111) slabs are composed of 13 atomic layers and the ensuing slab thickness is 27.44 Å. The three outermost atomic layers are relaxed, while the remaining atoms are fixed to their bulk positions. The calculated surface energy of Pd(111) is 0.091 eV/Å$^2$, in good agreement with reported theoretical values (0.082 eV/Å$^2$ by PBE[64], 0.099 eV/Å$^2$ by PBEsol[65]) and experiment[66] (0.125 eV/Å$^2$).

To account for the various coverages of B, H, and OH on the Pd(111) surface, different sizes of surface unit cells are employed. The coverage (Θ) of each adsorbate atom is defined as the ratio between the number of adsorbate atoms and the number of surface Pd atoms, with an equal number of respective atoms corresponding to 1 ML. For 0.25, 0.5 and 0.75 ML, a *p*(2×2) surface cell is used and for 1 ML a *p*(1×1) surface cell is used.

The binding energy ($\Delta E_{\mathrm{b}}$) is calculated as

$$\Delta E_{\mathrm{b}} = \frac{1}{2N_{\mathrm{ads}}}\left(E_{\mathrm{tot}}^{\mathrm{ads/subs}} - E_{\mathrm{tot}}^{\mathrm{subs}} - 2N_{\mathrm{ads}} \cdot \mu_{\mathrm{ads}}\right), \quad (Eq. 1)$$

where $E_{\mathrm{tot}}^{\mathrm{ads/subs}}$ and $E_{\mathrm{tot}}^{\mathrm{subs}}$ are the calculated DFT total energies of the adsorbate-substrate and substrate, respectively. $N_{\mathrm{ads}}$ is the number of adsorbates in the used surface supercell. $\mu_{\mathrm{ads}}$ is the chemical potential of adsorbates with respect to the relevant reference phases at the given experimental condition, defined as

$$\mu_{\mathrm{ads}}(c_{\mathrm{ads}}, T, pH) = E_{\mathrm{tot}}^{\mathrm{ads}} + \Delta\mu_{\mathrm{ads}}(c_{\mathrm{ads}}, T, pH)$$

where $E_{\mathrm{tot}}^{\mathrm{ads}}$ is the DFT calculated total energy of the adsorbates reference phases (*i.e.* the $H_2$ molecules for a H adsorbate and the rhombohedral α-phase of boron for a B adsorbate). $\Delta\mu_{\mathrm{ads}}$ is the chemical potential difference with respect to its standard reference phase where the source for the adsorbate atom should be an ion in solution. We note that $\Delta\mu_{\mathrm{H}}$ should be zero because the $H_2$ gas phase is the relevant reservoir for the here considered experiments, even in the electrochemical measurements. The source of the adsorbed B is an $\mathrm{BH}_4^-$ ion in solution, thus $\Delta\mu_{\mathrm{B}}$ is calculated using Ref. [67]as follows:

$$\Delta\mu_{\mathrm{B}}\left(c_{\mathrm{BH}_4^-}, T, pH\right) = \mu_{\mathrm{BH}_4^-} - 4 \cdot \mu_{\mathrm{H}} - \mu_e$$

where $\mu_{BH_4^-}$, $\mu_H$, and $\mu_e$ are the chemical potentials of $BH_4^-$ , H, and the electron, respectively. These can be calculated using tabulated literature data and the experimental conditions (*i.e.* concentrations of the ions, *p*H, and temperature) as described in the SI.

The *d*-band centre is calculated by averaging over the eigenenergy multiplied by the density of states (DOS) of occupied Pd-*d* states as specified by the formula:

$$E_{d-\mathrm{centre}} = \frac{\int_{-\infty}^{E_\mathrm{F}} E * \mathrm{DOS}(E)}{\int_{-\infty}^{E_\mathrm{F}} \mathrm{DOS}(E)}, \quad (Eq. 2)$$

where *E* and DOS(*E*) are the eigenenergy and the density of Pd-*d* states, respectively.

# Corresponding Author

Correspondence and requests for materials should be addressed to B.G. (email: b.gault@mpie.de) or S.-H.K. (email: s.kim@mpie.de)

# Author Contributions

S.-H.K. and B.G. designed the overall experiment. S.-H.K. performed the synthesis, co-electrodeposition, and atom probe specimen preparation. S.-H.K. measured APT and analysed the acquired data with support from B.G.. J.J. and J.L. performed HR-TEM, (S)TEM-EDS, and EELS measurements with support from C.S.. O.K. performed XPS and analysed data with A.E.-Z.. S.S. performed HOR test with support from O.K. and H.L.. S.-H.Y. performed DFT calculations with support from M.T. and J.N.. S.-H.K., S.-H.Y., M.T., and B.G. drafted the manuscript. All authors contributed and have given approval to the final version of the manuscript.

## Acknowledgements

We thank Uwe Tezins, Christian Broß, and Andreas Sturm for their support to the FIB and APT facilities at MPIE and Benjamin Breitbach for the XRD measurement. S.-H.K. and B.G. gratefully thank Dr. Leigh T. Stephenson and Dr. Kevin Schweinar for atom probe data analysis. S.-H.K., A.E.-Z., L.T.S., and B.G. acknowledge financial support from the ERC-CoG-SHINE-771602. S.-H.Y., J.N., and M.T. acknowledge support from the RESOLV program by the Deutsche Forschungsgemeinschaft (DFG, German Research Foundation) under Germany's Excellence Strategy-EXC 2033- 390677874-RESOLV and funding by the DFG through SFB1394, project no. 409476157.

# Controlled doping of electrocatalysts through engineering impurities

*Se-Ho Kim[1,†,*], Su-Hyun Yoo[1,†], Sangyong Shin[2], Ayman A. El-Zoka[1], Olga Kasian[1,3], Joohyun Lim[1,4], Jiwon Jeong[1], Christina Scheu[1], Jörg Neugebauer[1], Hyunjoo Lee[2], Mira Todorova[1], Baptiste Gault[1,5*]*

[1]Max-Planck-Institut für Eisenforschung GmbH, Max-Planck-Straße 1, 40237 Düsseldorf, Germany

[2]Department of Chemical and Biomolecular Engineering, Korea Advanced Institute of Science and Technology, Daejeon 34141, South Korea

[3]Helmholtz-Zentrum Berlin GmbH, Helmholtz Institut Erlangen-Nürnberg, 14109 Berlin, Germany

[4]Department of Chemistry, Kangwon National University, Chuncheon 24342, Republic of Korea

[5]Department of Materials, Royal School of Mines, Imperial College, London, SW7 2AZ, United Kingdom

[†] these authors contributed equally.

[*] co-corresponding authors.

# Contents

# Experimental Results

X-ray Photoelectron Spectroscopy (XPS). XPS analysis of the Pd-B nano-materials was mainly aimed at probing the extent of surface enrichment in B, and the presence of any other species. The spectra analyzed, (Figures S1 and S2) showed no indication of B or possible boron compounds in both samples, albeit a shoulder peak observed in Pd-$B_{40}$ at 190 eV might be due to B[1]. Worthy of note, however, is the consistent observation of possible Cl 2*p* peak at 196 eV, probably due to the $K_2PdCl_4$ reagent used in synthesis, peak in different samples.

The presence of B on the surface of Pd-B nano-materials is very scarce, suggesting it is embedded in the subsurface. The addition of B does seem to affect the oxidation state of the Pd on the surface, as the binding energies for Pd $3d_{3/2}$ and Pd $3d_{5/2}$ peak increasingly shift towards lower values with increased B doping, when it would be expected that the peak values would shift positively due to the feature size effect[2]. Furthermore, the deconvolution of the asymmetric Pd $3p_{3/2}$ and Pd $3d_{5/2}$ peaks (Figure S3) showed the existence of an oxide layer at the surface[3]. This was seen in previous XPS analyses of Pd nanoparticles[4], and since a separate O peak is not detected, this oxide layer might be less than a monolayer, or due to remaining oxygen in the chamber rather than a full PdO layer[5] formed during synthesis.

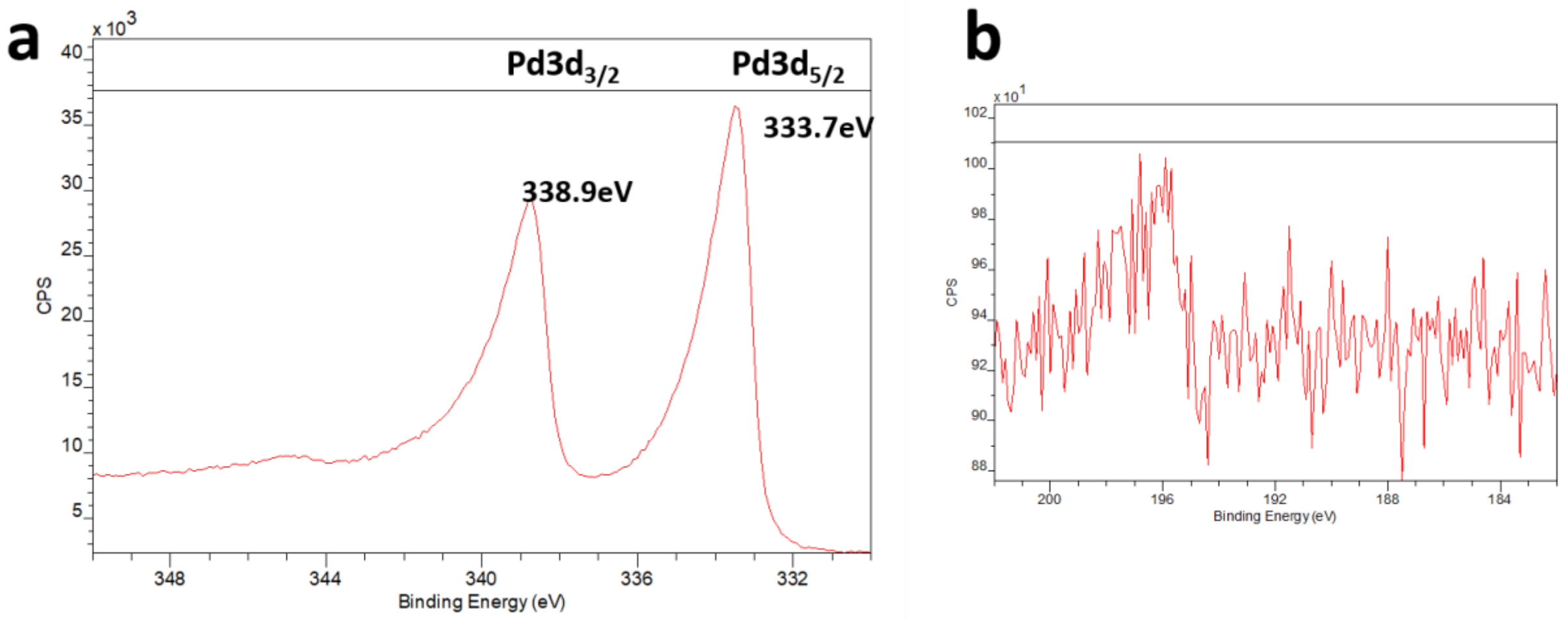

**Figure S1.** (a) Pd 3*d* and (b) B 1*s* XPS spectra of Pd-B1 sample. Pd 3$d_{5/2}$ and 3$d_{3/2}$ binding energies are 333.7 and 338.9 eV, respectively. A shoulder peak at 196 eV could correspond to possible Cl 2*p* (no B 1*s* related peak detected).

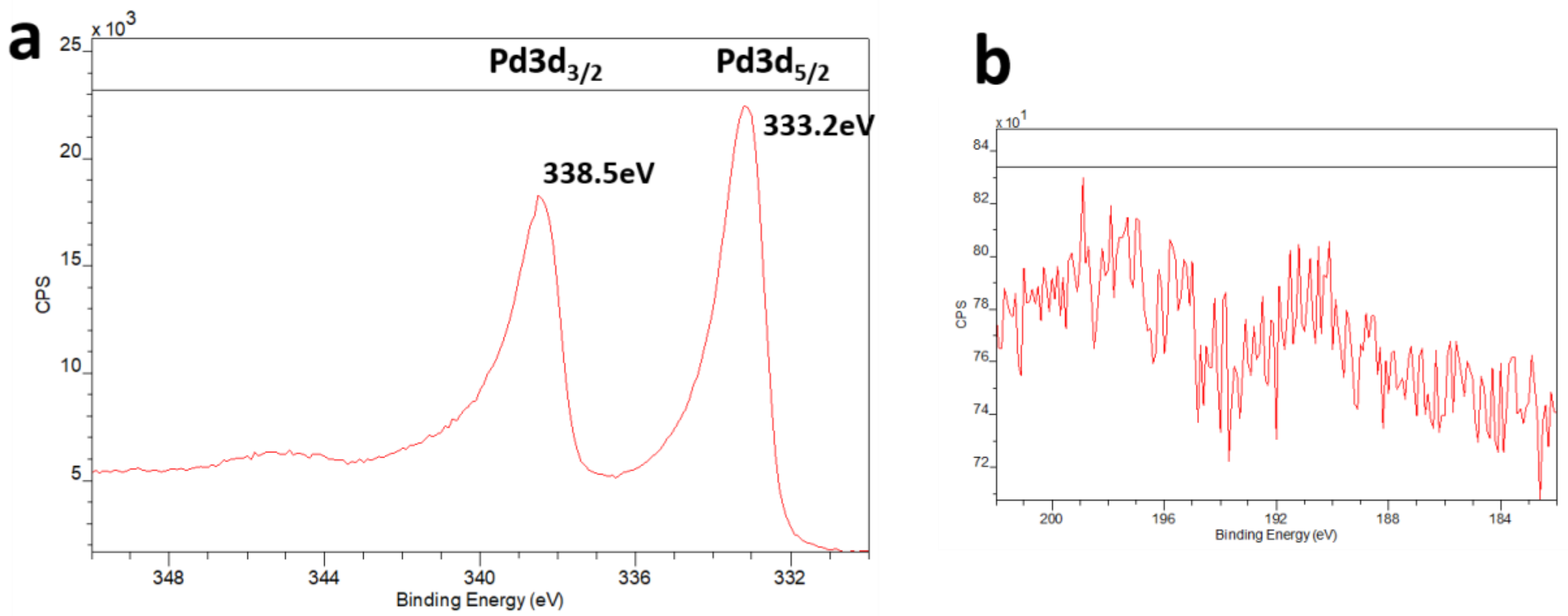

**Figure S2.** (a) Pd 3*d* and (b) B 1*s* XPS spectra of Pd-$B_{40}$ sample. Pd 3$d_{5/2}$ and 3$d_{3/2}$ binding energies are 333.2 and 338.5 eV, respectively. A shoulder peak around at 190 eV could correspond to B 1*s*.

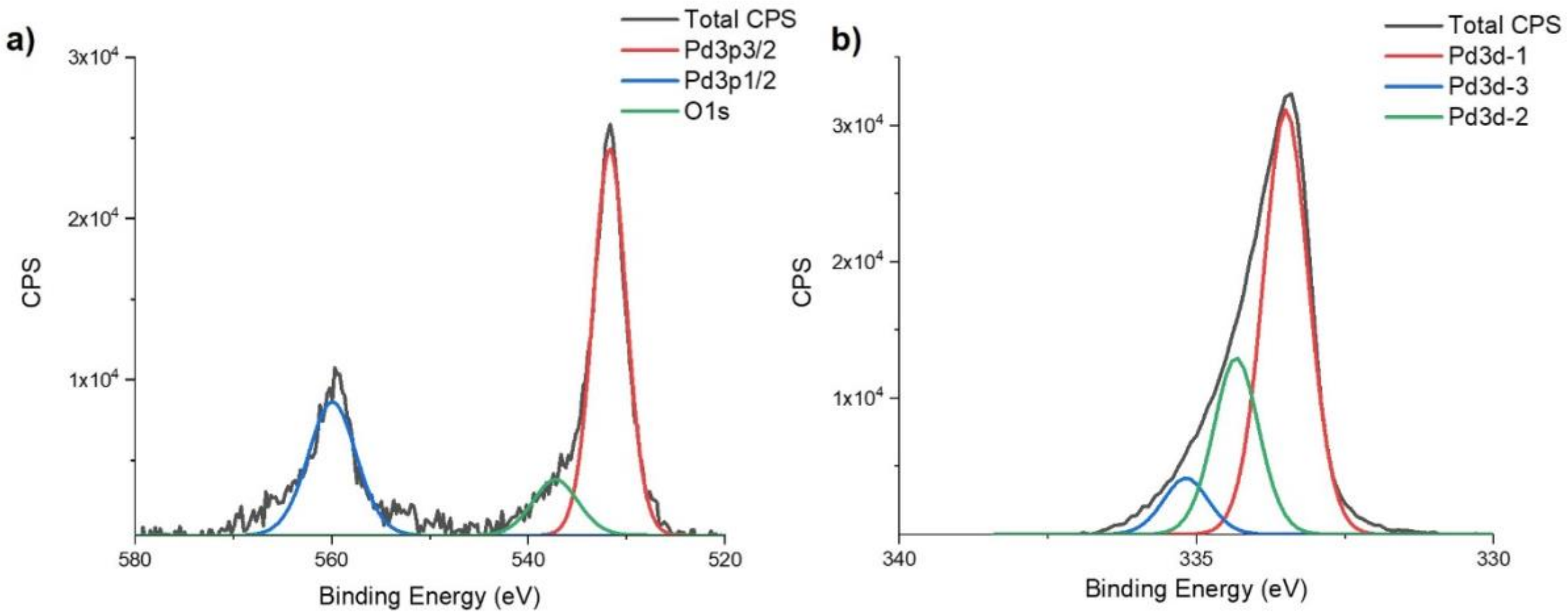

**Figure S3.** (a) Deconvolutions of the Pd 3*p*3/2 peak which coincides with the O 1*s* peaks. (b) deconvolution of the Pd 3*d*5/2 peak showing different states involved in the asymmetric peak detected. Pd 3*d*-1 is attributed to surface Pd, Pd 3*d*-2 is attributed to bulk Pd, and Pd 3*d*-3 can be attributed to the sublayer of an oxide.

**Aerogel-like Pd-$B_x$ nanoparticle synthesis.** Although nanoparticles synthesized with $NaBH_4$ reduction method are commonly accepted as a pristine nanoparticle, chemisorption of $BH_4^-$ from $NaBH_4$ into a metal (M) site[6,7] could produce unstable intermediate species[8]: M-$BH_3$. There is a possibility that M-$BH_3$ intermediate species could further react with early stages of nanoparticles (atoms, nuclei) and decompose into B atoms of which it is possible to incorporation inside nanoparticle system during nucleation and growth process[9–11] (see the Pd seed synthesis section for details). Moreover, B is thermodynamically favorable to dissolve on interstitial sites of metals[12,13].

To originate every species from each characterization tools, we chose a simple Pd nanomaterials system synthesized without any surfactant for our models. Since we did not add any C-based chemicals during Pd nano-aerogel synthesis, we expect all detected C-based species are impurities from the sampling process. The nanomaterials synthesized at different mole ratios of Pd precursor ($K_2PdCl_4$) to reductant ($NaBH_4$) at 0.1, 1, and 40 are named in Pd-$B_{0.1}$, Pd-$B_1$, and Pd-$B_{40}$, as described in Table S1. Then each Pd nanoparticle is investigated and compared to show consequences of B in terms of chemistry, morphology, and properties.

**Table S1.** Pd nano-materials synthesis with different molar ratios between Pd precursor and reductant.

| | **Pd-$B_{0.1}$** | **Pd-$B_1$** | **Pd-$B_{40}$** |
|---|---|---|---|
| [$K_2PdCl_4$], M | 0.01 | 0.01 | 0.01 |
| [$NaBH_4$], M | 0.001 | 0.01 | 0.4 |
| Ratio | 0.1 | 1 | 40 |

*Note that Pd-$B_1$ mole ratio of one is the ratio that used for a common nanoparticle seed synthesis.

**Sample preparation for APT measurement.** As-synthesized Pd nano-materials were prepared into APT sample following modified co-electrodeposition technique[14,15]. A common Ni ion electrolyte uses boric acid for a buffer acid; however, our goal in this research is to locate B atoms. To be confident that detected B atoms are originated from $NaBH_4$, citric acid was used to replace boric acid for Ni ion electrolyte. First, 15 g of nickel (II) sulfate hexahydrate (98%, Sigma-Aldrich) and 2.25 g of citric acid (99.5%, Sigma-Aldrich) were dissolved in 50 mL of distilled water. As-synthesized Pd nano-materials were then dispersed in the prepared Ni ions solution using a sonicator for 20 min followed by pouring the solution on a vertical cell. The cell includes a Cu substrate and a Pt counter electrode. Semi-bright Ni film was obtained by setting the constant current of -38 mA. Figure S4 shows scanning electron microscopy (SEM) images of the Ni-citric acid (Ni-cit) sample's surface at different magnifications. To optimize electroplating thickness for APT specimen preparation from Ni-cit electrolyte, different electrodeposition times were set at 100, 250, 500, and 1000 sec (see Figure S5). Cross-sectional images from focused-ion beam (FIB)-SEM show that the thickness increases as deposition time increases (a linear correlation plot is presented in Figure S6). We chose 1250 sec for Ni-cit electroplating time for complete encapsulation of Pd nanoparticles within Ni.

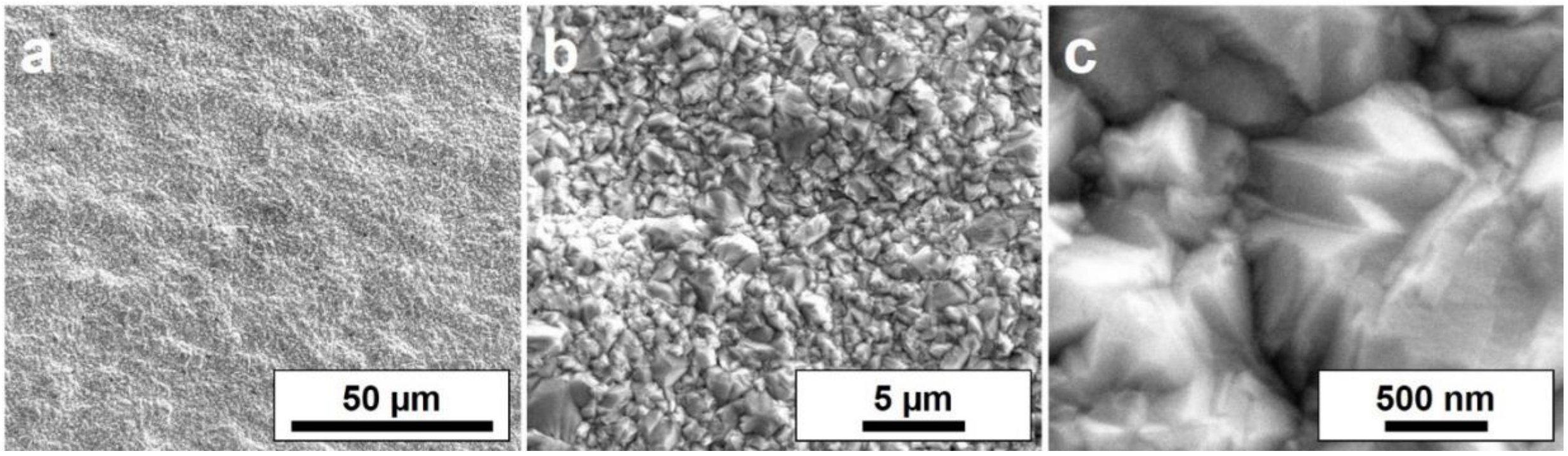

**Figure S4.** FIB-SEM surface images of electroplated Ni using citric acid as a buffer acid: (a) x250, (b) x2500, and (c) x25000 magnifications.

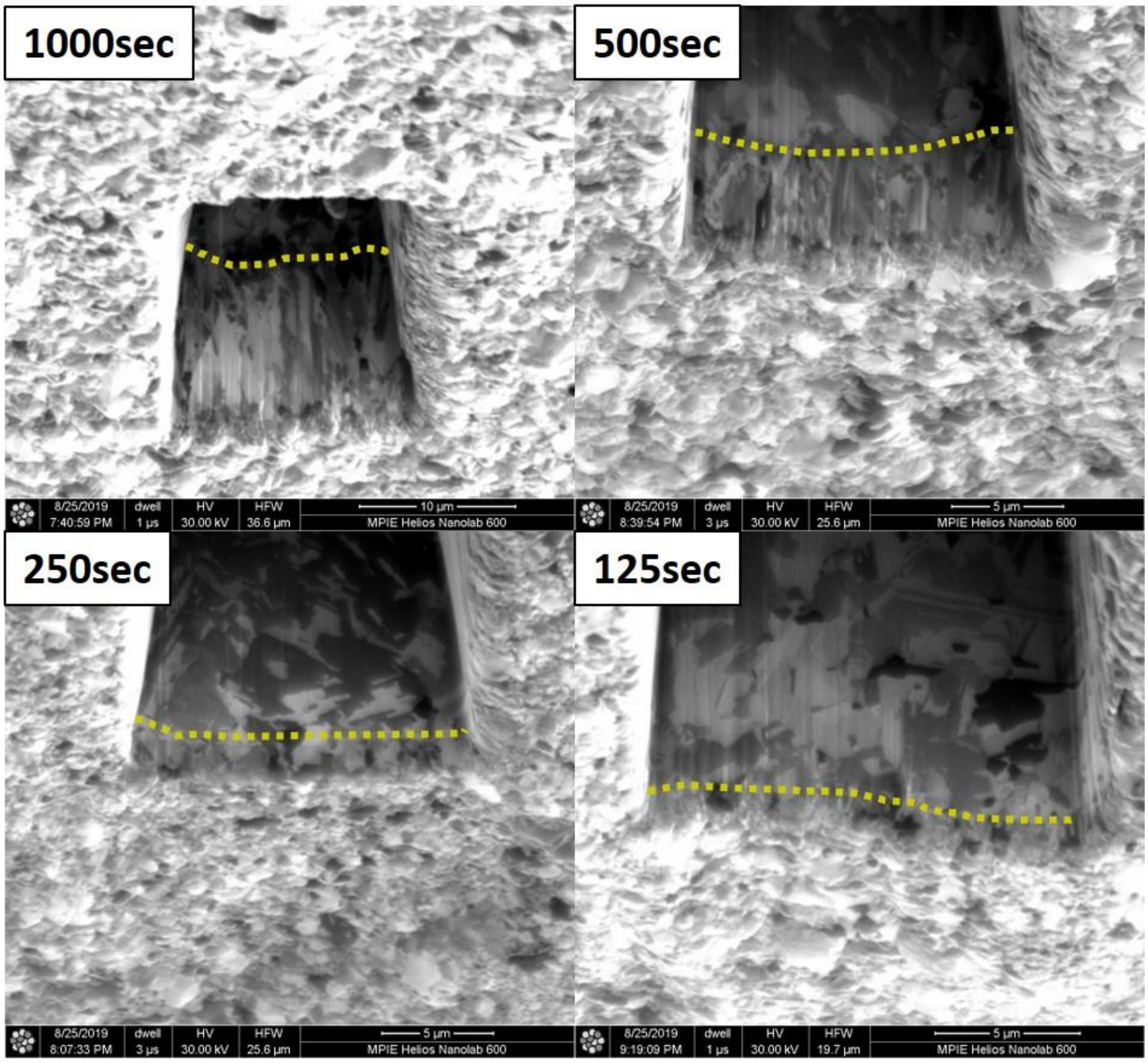

**Figure S5.** FIB cross-sectional ion-beam images of electroplated Ni at different deposition time: 1000, 500, 250, and 100 sec. All Ni films were electro-deposited at the constant current of -38 mA. Yellow dotted lines represent interfaces between electroplated Ni and Cu substrate.

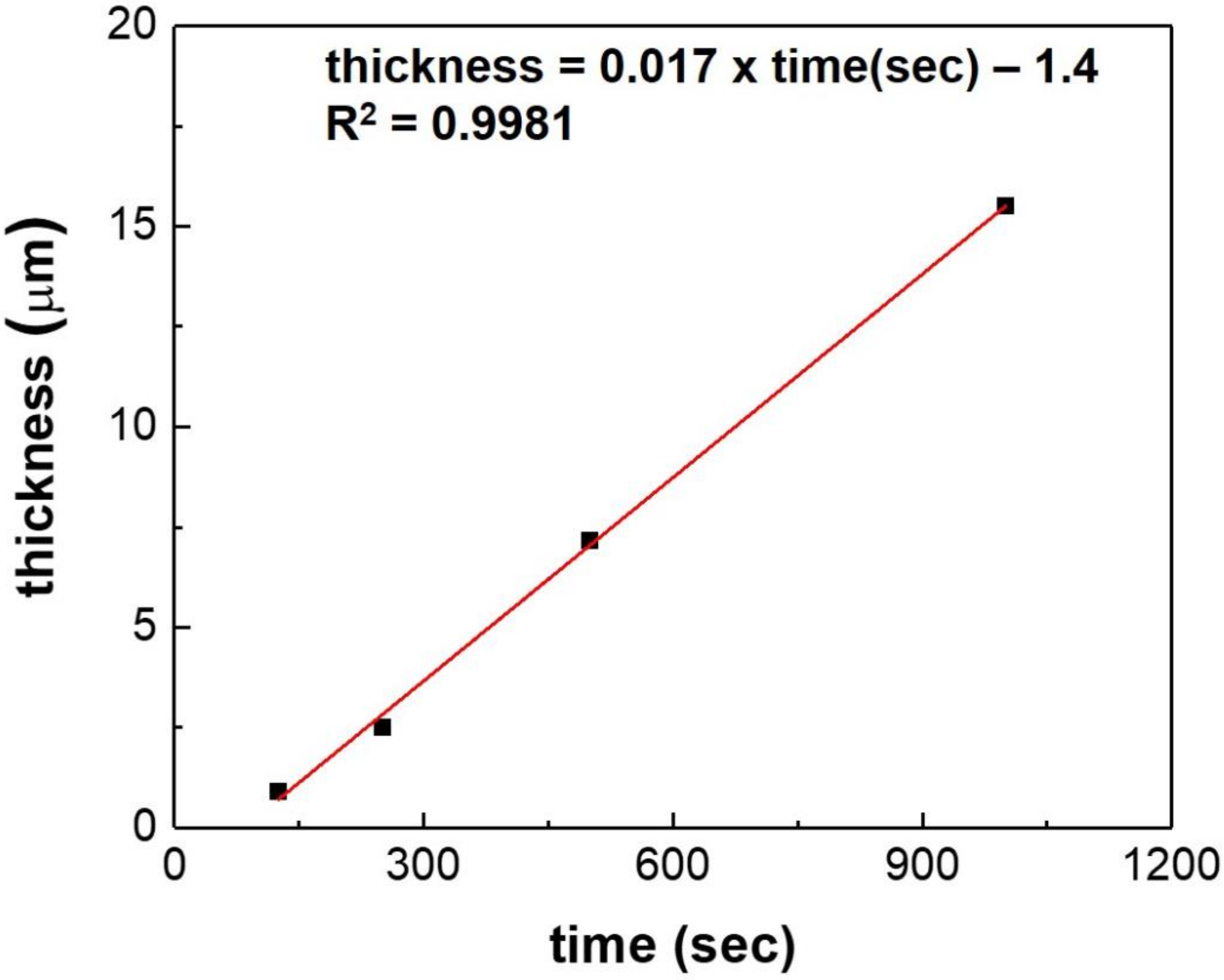

**Figure S6.** Ni film thickness *versus* deposition time from Ni-citric acid electrolyte at a constant current of -38 mA.

Encapsulation aerogel Pd nanoparticles in Ni layer. After electrodeposition, APT specimen was fabricated using FIB. On the composite Ni surface, there are protrusions as shown in Figure S7a, which indicates the presence of embedded nanoparticles[15]. The aggregated Pd nanoparticles inside Ni are confirmed by cross-sectioning one of protrusions (Figure S7b). We found no noticeable voids within Pd/Ni interfaces; therefore, these regions were lifted out and sharpened into a needle-like APT specimen as shown in Figure S7c.

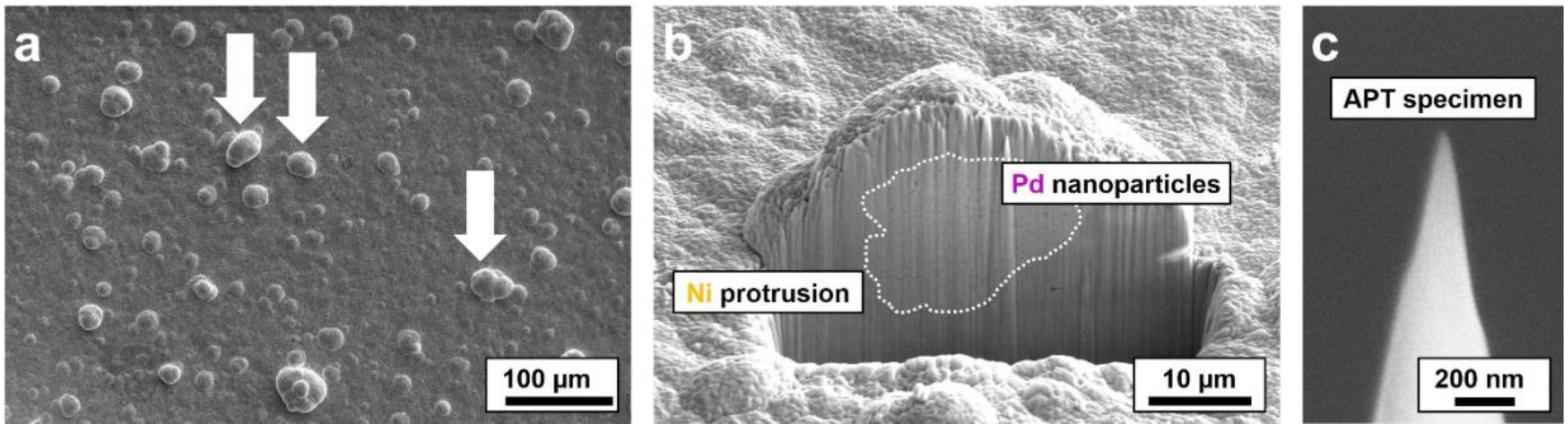

**Figure S7.** FIB-SEM (a) surface image of co-electrodeposited Pd nanoparticle/Ni sample with protrusion regions (white arrows) and (b) cross-sectional image of one of protrusions. (c) A sharpened specimen from Pd nanomaterials/Ni sample for APT measurement.

APT measurement. Ni is selected for a suitable electroplating matrix. Since it has a similar ion evaporation field with Pd (Ni = 35 V $nm^{-1}$, Pd = 37 V $nm^{-1}$)[16,17], trajectory artifacts would be minimal during field ion desorption process[18]. All reconstructed 3D atom map does not exhibit any strong density (see inset in Figure S8) nor severe voltage fluctuation (Figure S8) indicating that field evaporation of elements is homogenous and thus ensured good data integrity. Moreover, Ni is composed of five stable isotopes (58 to 61 Da) that do not overlap with any isotopes of Pd (102 to 110 Da) resulting precise composition analysis.

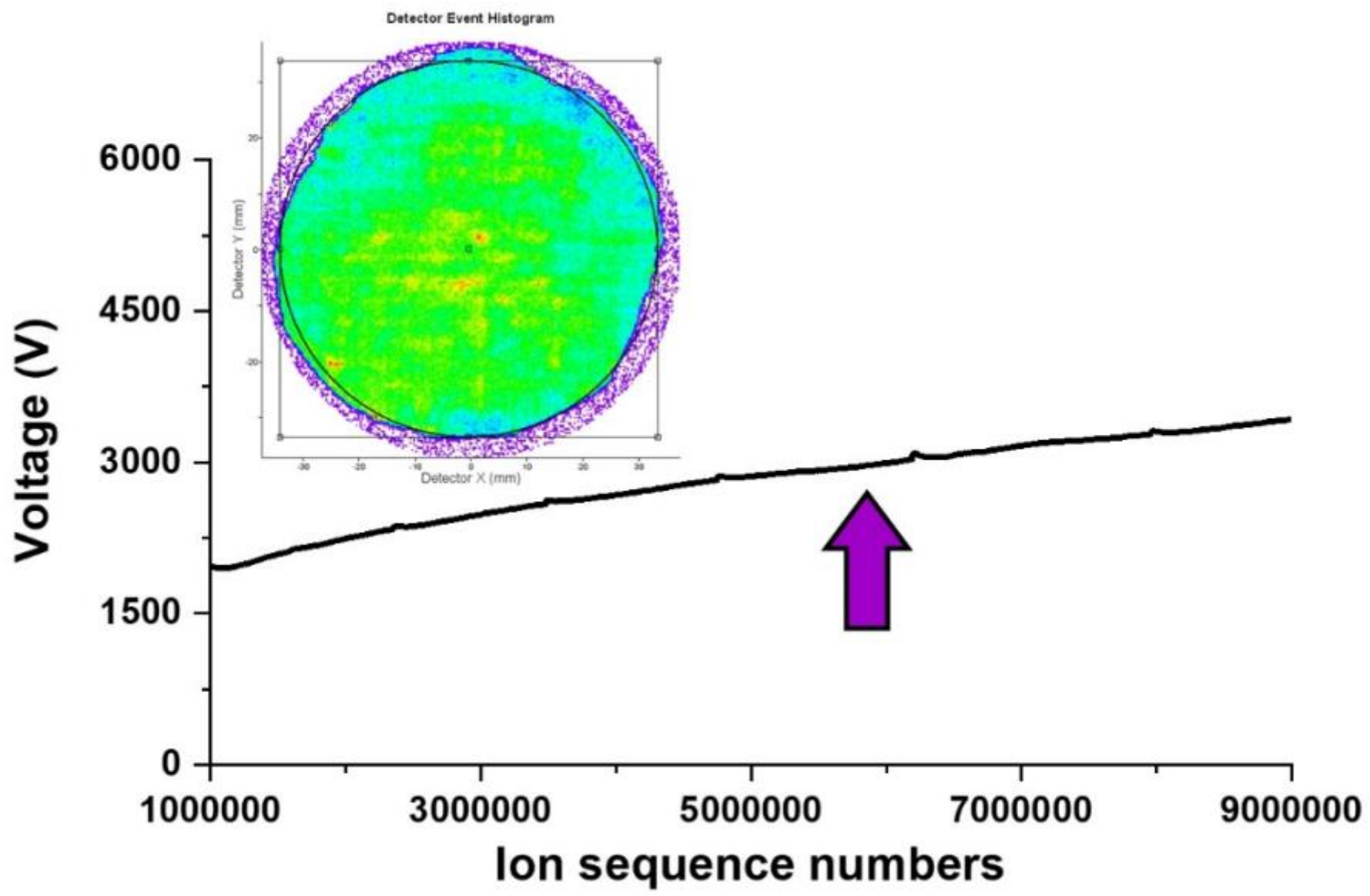

**Figure S8.** An example of voltage history curve of Pd-B in Ni measurement. Note that there is no voltage drop/increase when embedded Pd-B nanomaterials start to measure with Ni (marked by the purple arrow). Inset shows the detector event histogram of overall ions.

Mass spectrum analysis from acquired APT data. Figure S9 show examples of mass spectra of Pd-$B_{0.1}$ nano-catalysts embedded in Ni matrix, respectively, from the acquired APT data. Pd atoms are detected with Ni atoms. B is also detected at 10 and 11 Da which correspond to the natural isotopes of B for both samples. $C^+$, $C_2^+$, and $CO_2^+$ ions originated from citric acid are also detected that citric acid deposited within Ni matrix during sample preparation process[19,20]. Indeed, in Figure S10, a mass spectrum of a Ni matrix without Pd nanoparticle also shows strong peaks of C-based molecular ions.

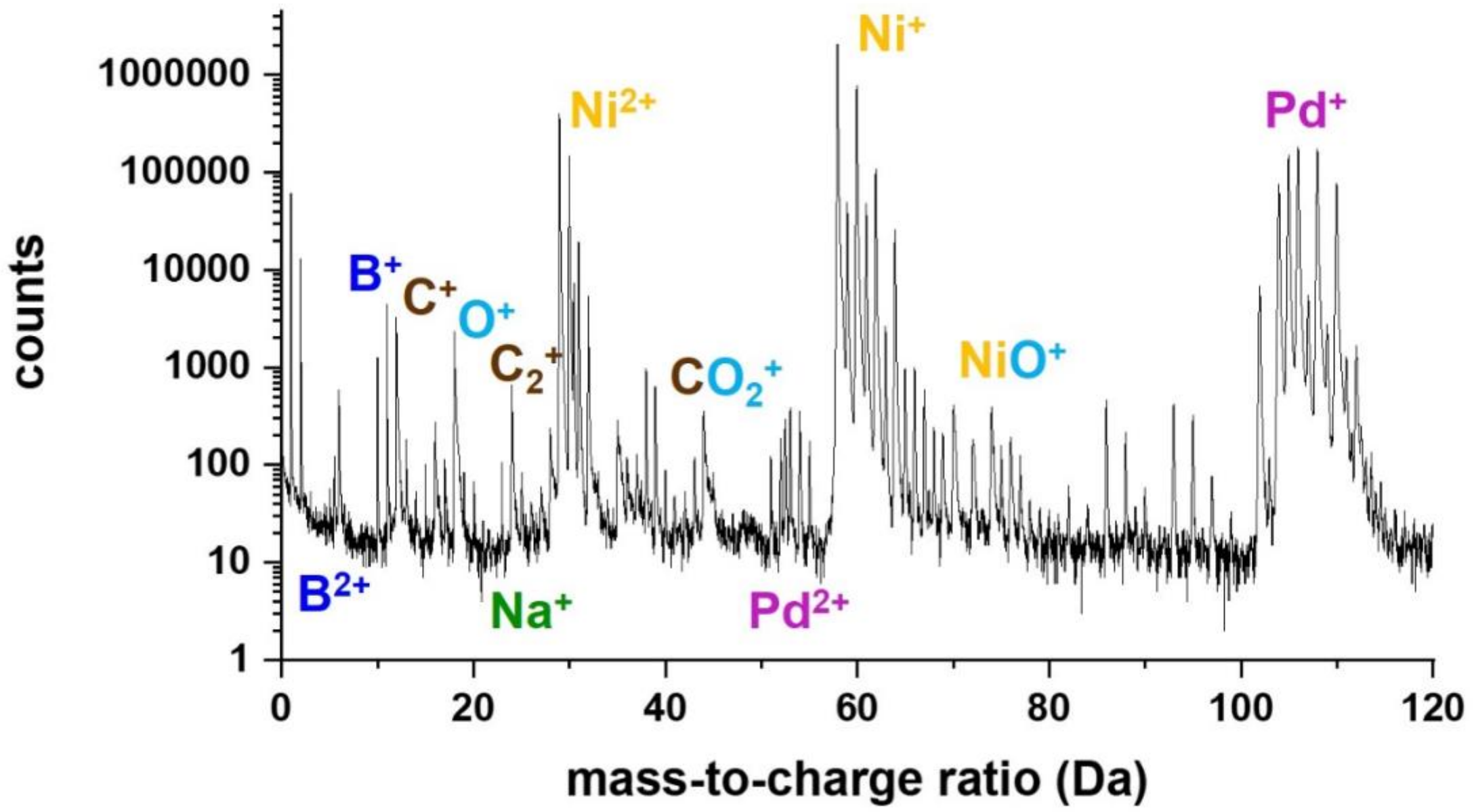

**Figure S9.** Mass spectrum of Pd-$B_{0.1}$ APT sample.

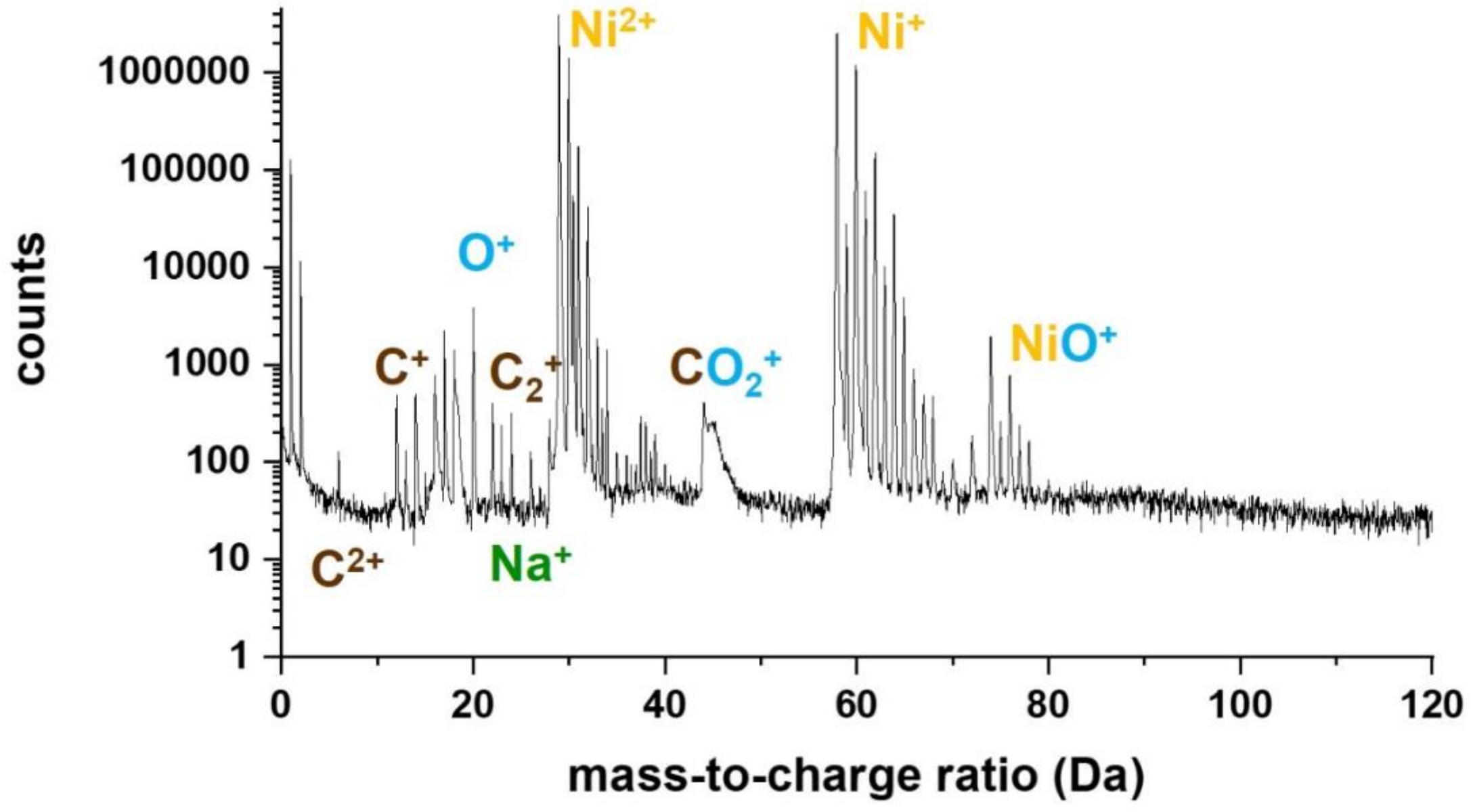

**Figure S10.** Mass spectrum of Ni matrix (no Pd nor B peaks detected).

Multi-events of Pd-B **nano-catalysts** in Ni sample. To investigate an ion loss during field evaporation, we plot the multiple-event correlation histograms on acquired in Figure S11. Evidence of dissociation nor neutral formation during field-evaporation is not observed in the acquired data which implies that it ensures a good data integrity. Figure S12 shows an example of B and Pd multiple-event histogram collected at the detector and the results supports that B atoms are located along with Pd atoms as they evaporate together.

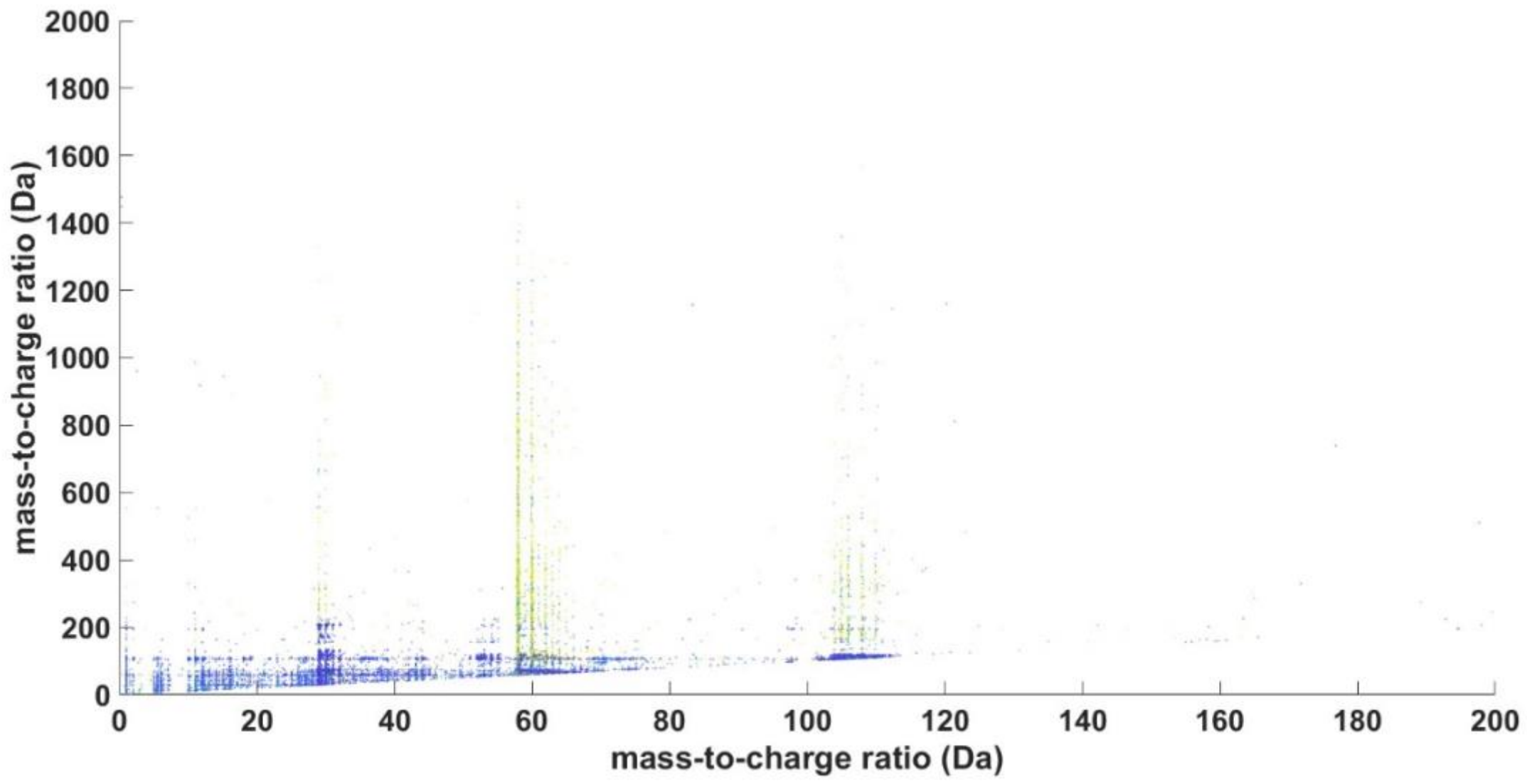

**Figure S11.** Correlation histograms of multi-events for Pd-B/Ni field evaporation. Note that there is no significant neutral formation or molecular ion dissociation trails.

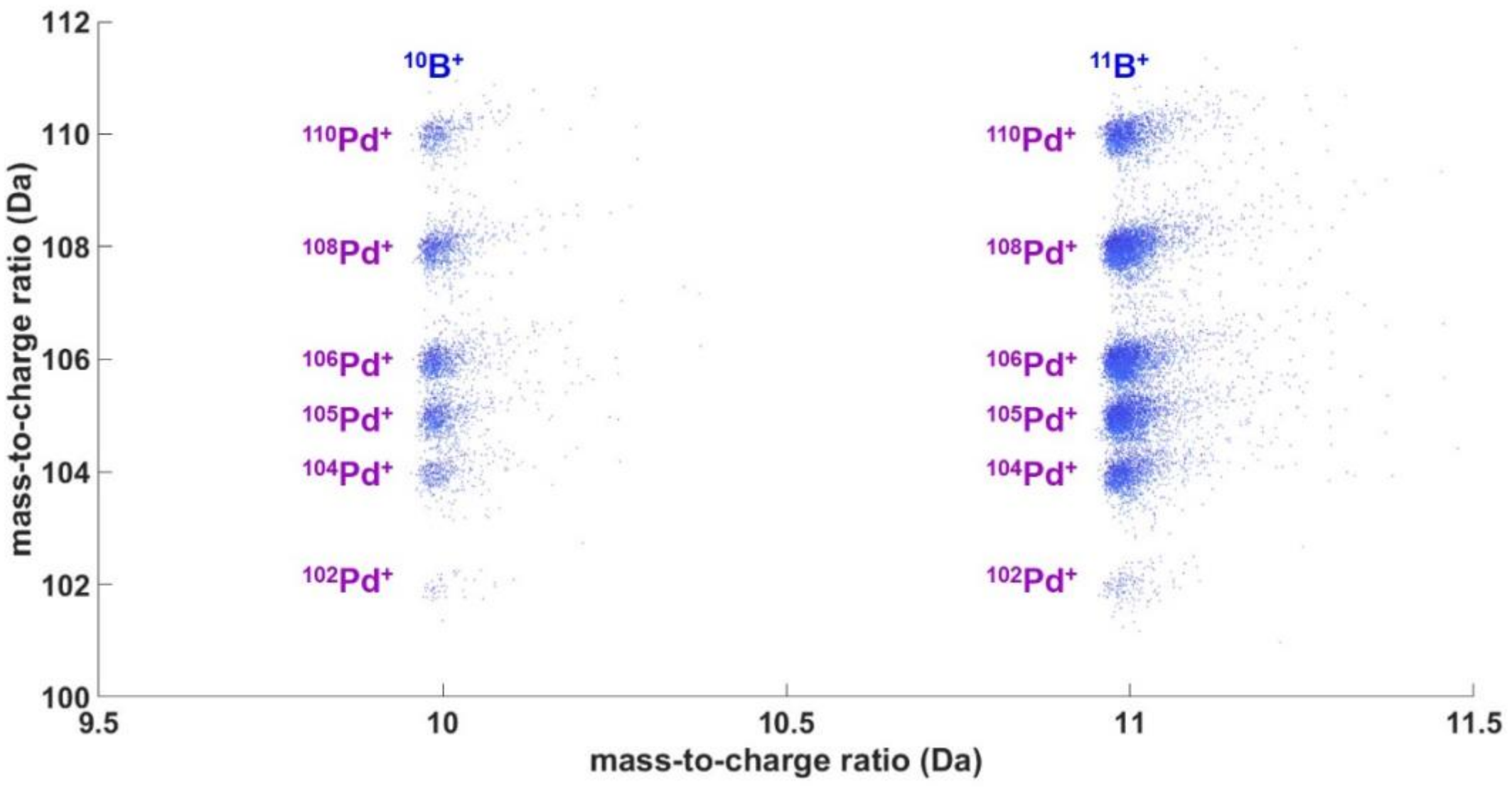

**Figure S12.** Mass-to-charge ratios for B (10, 11 Da) and Pd (102 - 110 Da) ions are shown to provide an evidence that they evaporate along each other. No dissociation trail is observed.

**B distribution on Pd-$B_{0.1-40}$ nano-materials.** To investigate whether B atoms are segregated or distributed homogeneously, we performed a B-B nearest-neighbor distance distribution analysis on each Pd-$B_x$ nanoparticle. The experimental result is compared with the simulated B atom distribution on same atomic positions. Figure 13a to 13c show that there is no significant deviation from each randomized/simulated result and the center of each B-B experimental curve ($x_c$) shifts to lower value as B concentration increase within the sample. For quantitative determination of the randomness, Pearson coefficient ($\mu_B$) of B (solute atom) was further measured associated to a $\chi^2$-statistical test. In short, $\mu_B = 0$ is equivalent to complete randomness while $\mu_B = 1$ means a clustered/ordered species. With a $\mu_B = 0.1275$, B is likely to be randomly distributed in the Pd-$B_{0.1}$ sample. The determined Pearson coefficient ($\mu_B$) of B (solute atom) are 0.1275, 0.1818, and 0.2090 for Pd-$B_{0.1}$, Pd-$B_1$, and Pd-$B_{40}$, respectively, which suggests that B species in Pd-$B_x$ nanoparticle system are homogenously distributed (Figure S13d).

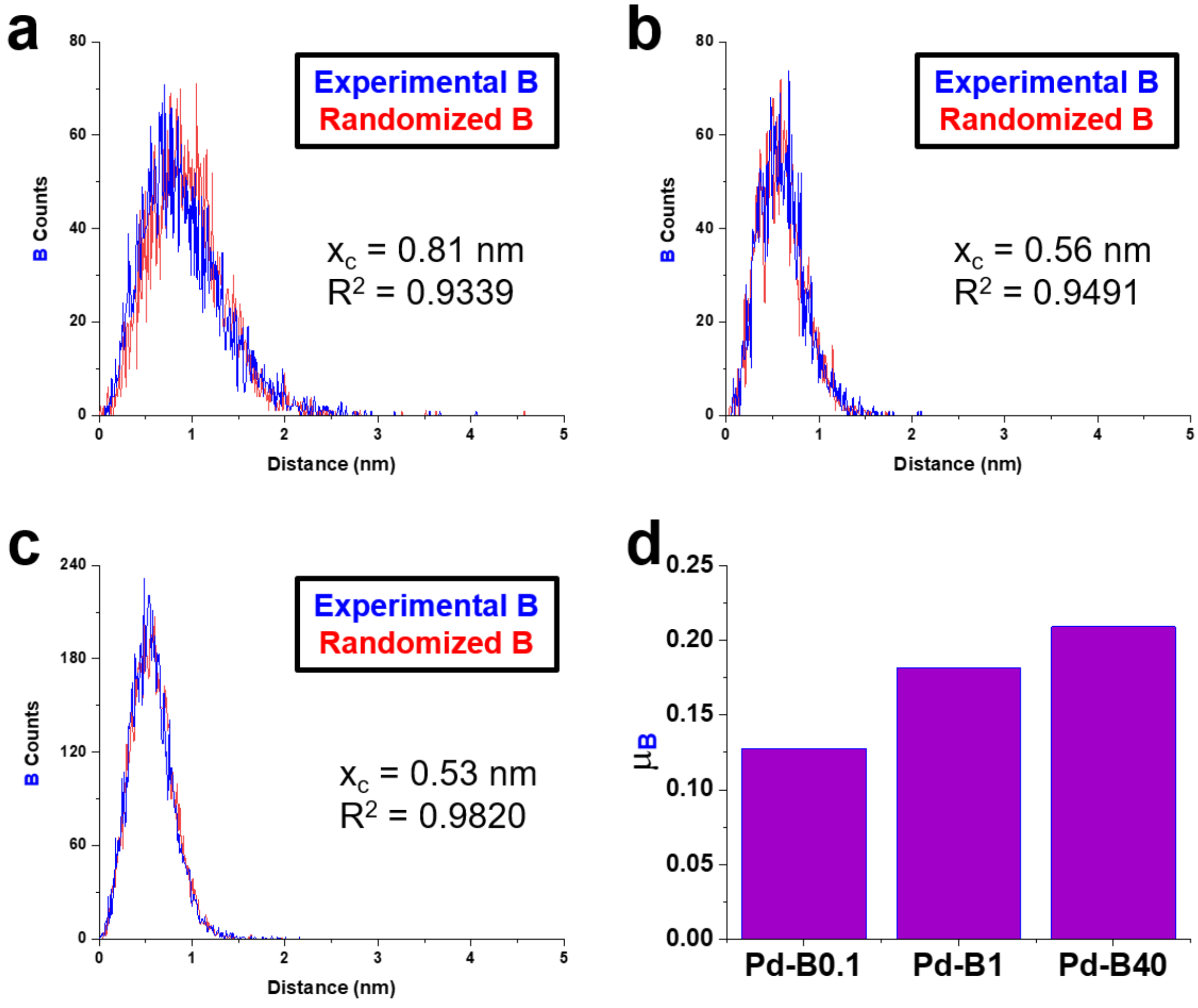

**Figure S13.** B-B nearest neighbour experimental (blue) and simulated (red) distance distribution of (a) Pd-$B_{0.1}$, (b) Pd-$B_1$, and (c) Pd-$B_{40}$. The center of B-B experimental distance distribution is calculated from a fitted Gaussian curve. (d) A comparison plot of the Pearson coefficients ($\mu_B$) of B on each sample.

**Mechanism for B incorporation in Pd nanoparticles.** Based on the calculated surface binding energy analysis, the calculated thermodynamic equilibrium concentrations of B in Pd bulk, and the experimentally tuned B concentrations in Pd nanoparticles, a mechanism for B incorporation in Pd nanoparticles is suggested in the following and visualized in Figure S14. Even though the binding energy for B adsorption in an on-surface hollow site is roughly 1 eV/(B atom) less favorable than the sub-surface octahedral site, B atoms may bind on the surface (Figure S14a), if there is a barrier hindering penetration into the sub-surface region. B atoms bound on the surface will be covered by an overgrowing Pd layer (Figure S14b), resulting in B atoms naturally occupying the more stable octahedral site (Figure S14c). The octahedral sites in the sub-surface layer and further away from the surface in the Pd bulk region are energetically favorable (*e.g.*, the formation energy of B at 0.25 ML in the sub-surface layer is -1.24 eV/B atom and -1.47 eV/B atom for 0.4 B at. % in Pd bulk, respectively). Therefore, B atoms integrated into Pd nanoparticles will naturally remain in the octahedral sites below the surface, as diffusion of B out of Pd-bulk/subsurface to rejoin the chemical reservoir is an endothermic reaction (Figure S14d).

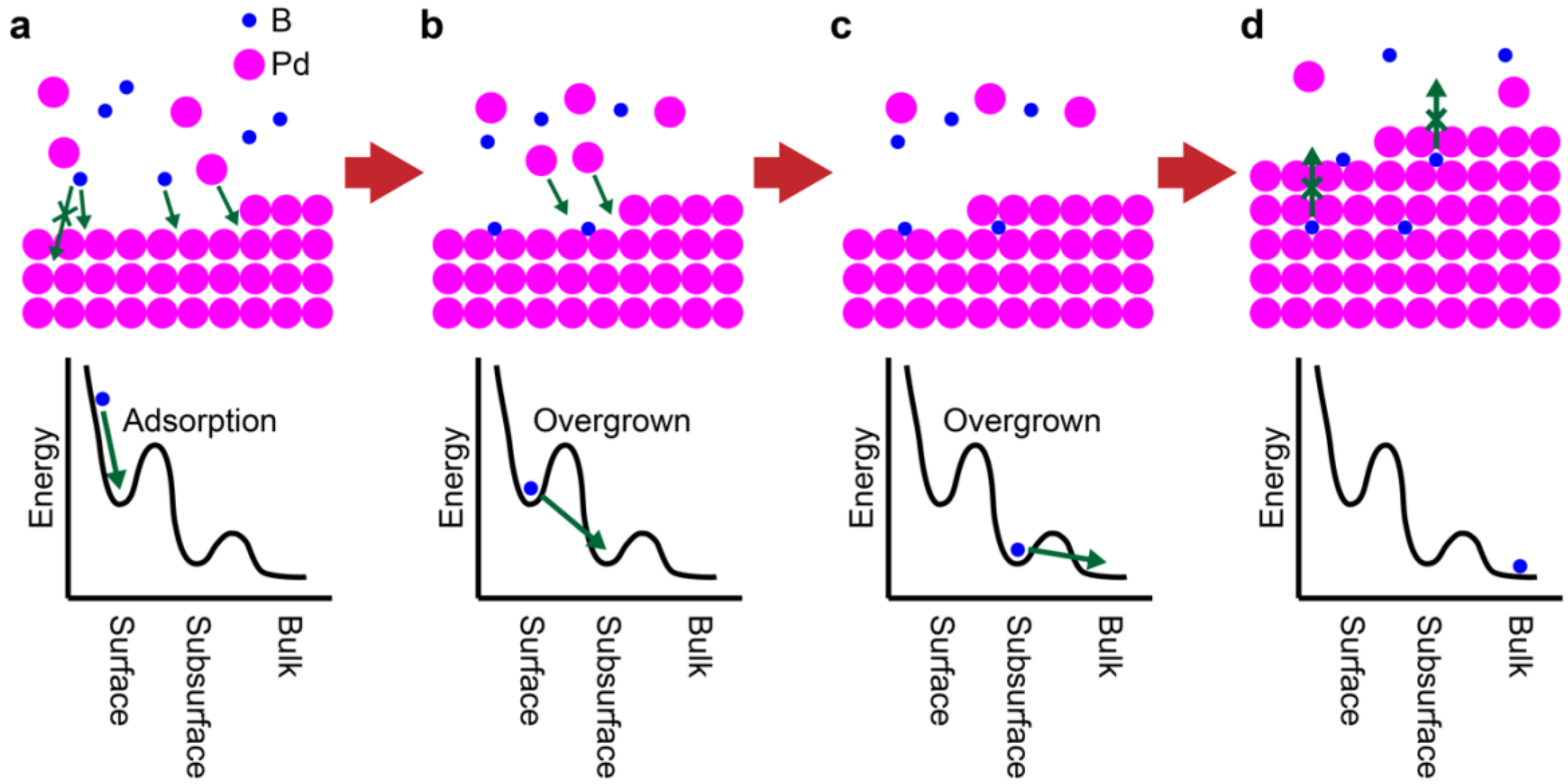

**Figure S14.** Schematic picture for the theoretically suggested mechanism of B incorporation in Pd nanoparticles. The upper panels show the atomic configurations of incorporating B and Pd atoms from chemical reservoirs into the surface of a Pd nanoparticle for different stages during growth, while the bottom panels show the corresponding energy landscape. Purple and blue circles show Pd and B atoms, respectively. Green arrows with a cross indicate that diffusion along the corresponding direction is negligible.

B introduction from hydrogenation of $NaBH_4$. A Pd wire ($\Phi$ 0.25 x 18 $mm^3$, 99.99+% (Goodfellow)) is immersed into the excess amount of $NaBH_4$ solution (1.0 M) as shown in Figure S15a. The calculated mol ratio of B atoms in 1.0 M of $NaBH_4$ to Pd atoms in the wire (density = 12.02 g $cm^{-3}$, atomic mass = 106 g $mol^{-1}$) is ~100. After bubbling stopped completely, the wire is rinsed with distilled water once and dried. After an hour, it is mounted on a FIB holder for APT specimen preparation (Figure S15b). A final needle-like APT specimen is shown in Figure S15c. To investigate whether B can be introduced after the synthesis, APT measurement is performed for the post $NaBH_4$-treated Pd wire sample. Although 100 times more moles of B source is used, the mass spectrum analysis shows trace amount of B and overall composition analysis shows that there is 0.001 at.% of B in Pd wire (Figure S16). 3D atom map shows that B is distributed randomly and no sign of clustering is observed. Therefore, B in the Pd nano-materials can be introduced more likely from the synthesis not from post-treatment using $NaBH_4$.

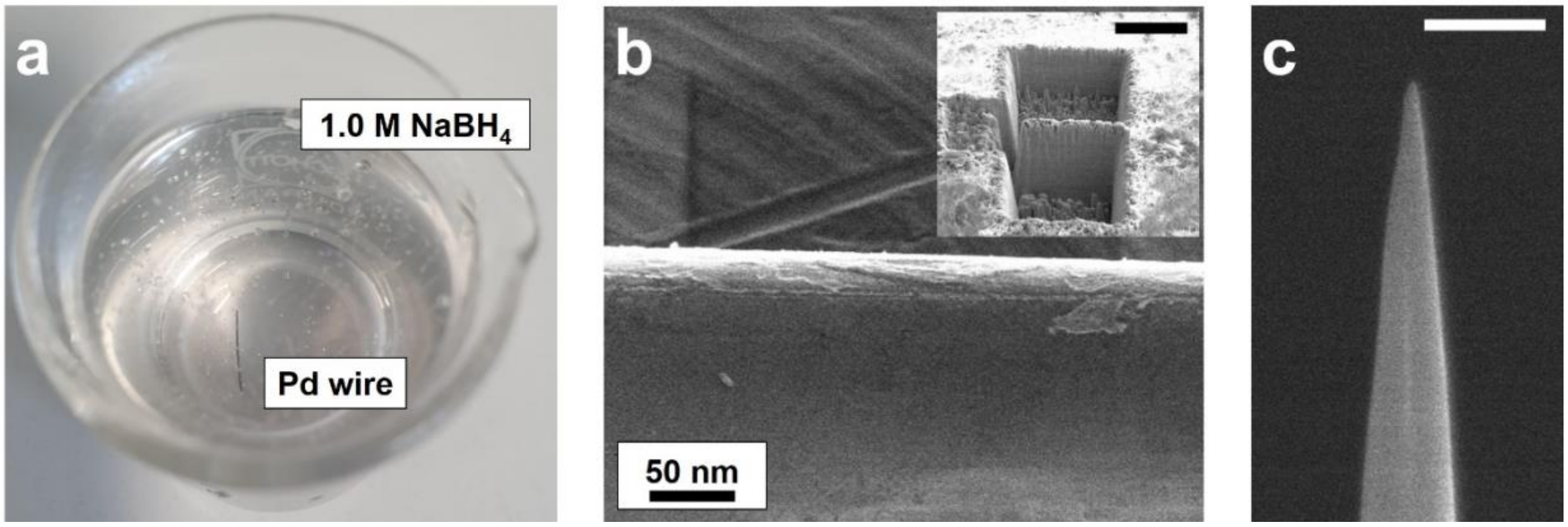

**Figure S15.** (a) Pd wire is immersed in 1.0 M $NaBH_4$ solution until $H_{2(g)}$ generation stopped. (b) FIB-SEM image (tilted at 52$^o$) of the as-reacted Pd wire. Inset shows the cross-sectional image before the lift-out process performed. (c) A final APT specimen of the Pd wire.

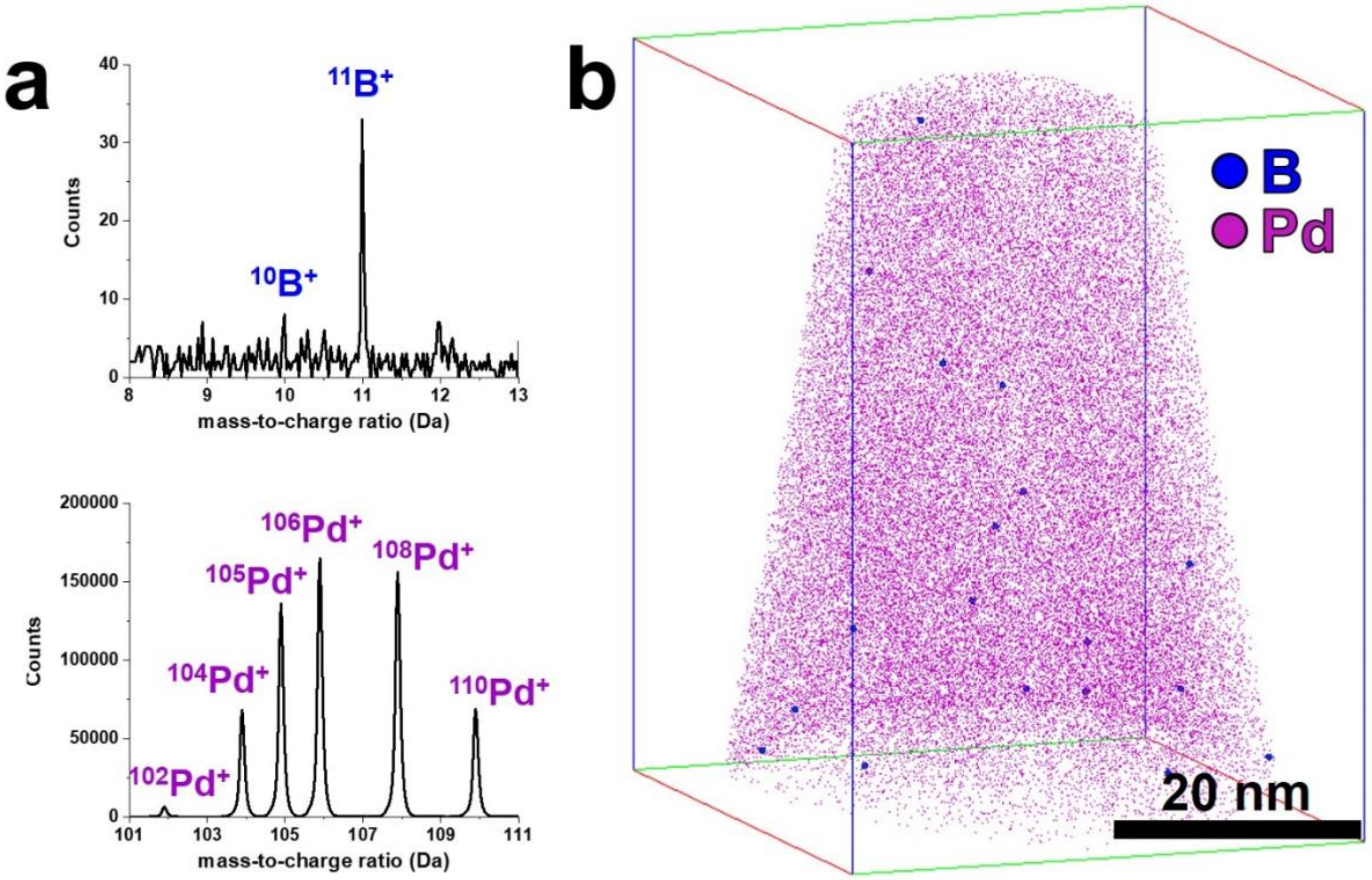

**Figure S16.** (a) Sectioned mass spectra of $B^+$ and $Pd^+$ ions ranges. (b) 3D reconstruction of the Pd wire. Pd and B atoms are colored in purple and blue, respectively.

**Pd-$B_1$ and Pd-$B_{40}$ nano-catalysts TEM analysis.** Figure S17a-c show (S)TEM images of the as-synthesized Pd gels of Pd-$B_{0.1}$, Pd-$B_1$, and Pd-$B_{40}$, respectively. All nanoparticles are aggregated forming into a complex aerogel-type structure and this is attributed to the absence of surfactant which is often used to stabilize nanoparticle surface and avoid agglomeration. High-resolution (HR)-TEM images of Pd-$B_{0.1}$, Pd-$B_1$, and Pd-$B_{40}$ in Figure S17d-f, respectively, show all has a poly-crystalline structure. Chen et al. reported that the structural distortion by B doping in face centered cubic (FCC) lattice can locally lead to a small phase transformation to hexagonal close packed (HCP) structure[21,22]. Here, the powder XRD results in all sample show that a dominate phase is FCC and also in the HR-TEM.

In the classical nucleation and growth model of nanoparticles, it proposed that reduced atoms rapidly aggregate to form nuclei[23,24]. These nuclei are then further grown to nano-crystalline particles. A few regions in Pd-$B_{40}$ have multiple twinned nanocrystal (decahedron), which indicates outburst generation of atoms and nuclei[25]. A decahedron particle is reported as an initial stage of nano-crystalline and these structures are observed within the aerogel structure[26] (see Figure S18).

To investigate any presence of B, EDS-STEM analysis was performed and the B signal was detected in the Pd nano-materials system (see Figure S19). However, the detailed information such as composition and element profile were difficult to be obtained due to its relatively less excited X-ray signals production from the accelerated electrons[27].

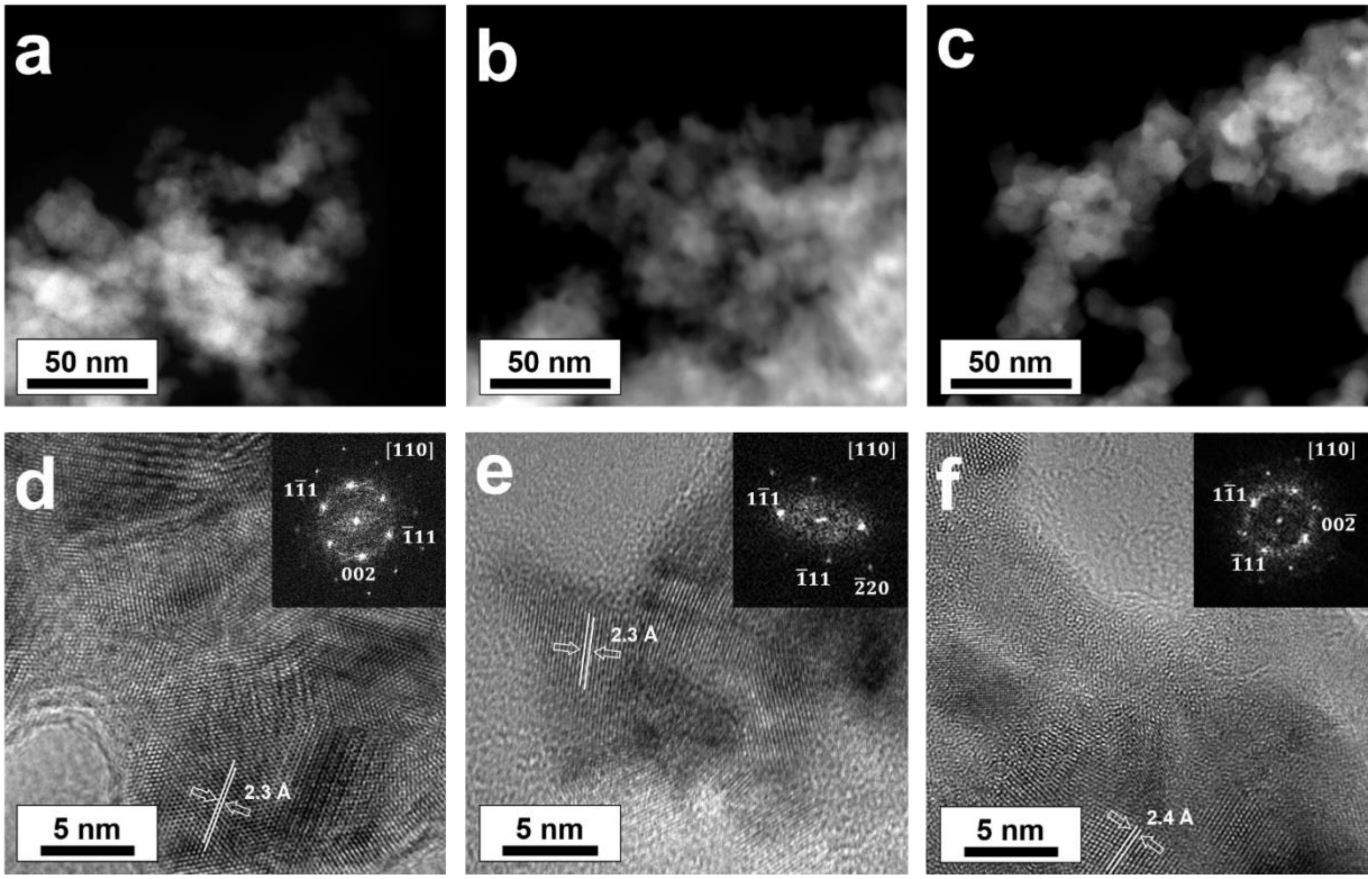

**Figure S17.** (S)TEM and HR-TEM images of (a,d) Pd-$B_{0.1}$, (b,e) Pd-$B_1$, and (c,f) Pd-$B_{40}$ nano-materials, respectively. Inset images display corresponding Fast Fourier Transformation (FFT) patterns on nano-grains where d-spacings are measured along [110] $Pd_{fcc}$ zone axes.

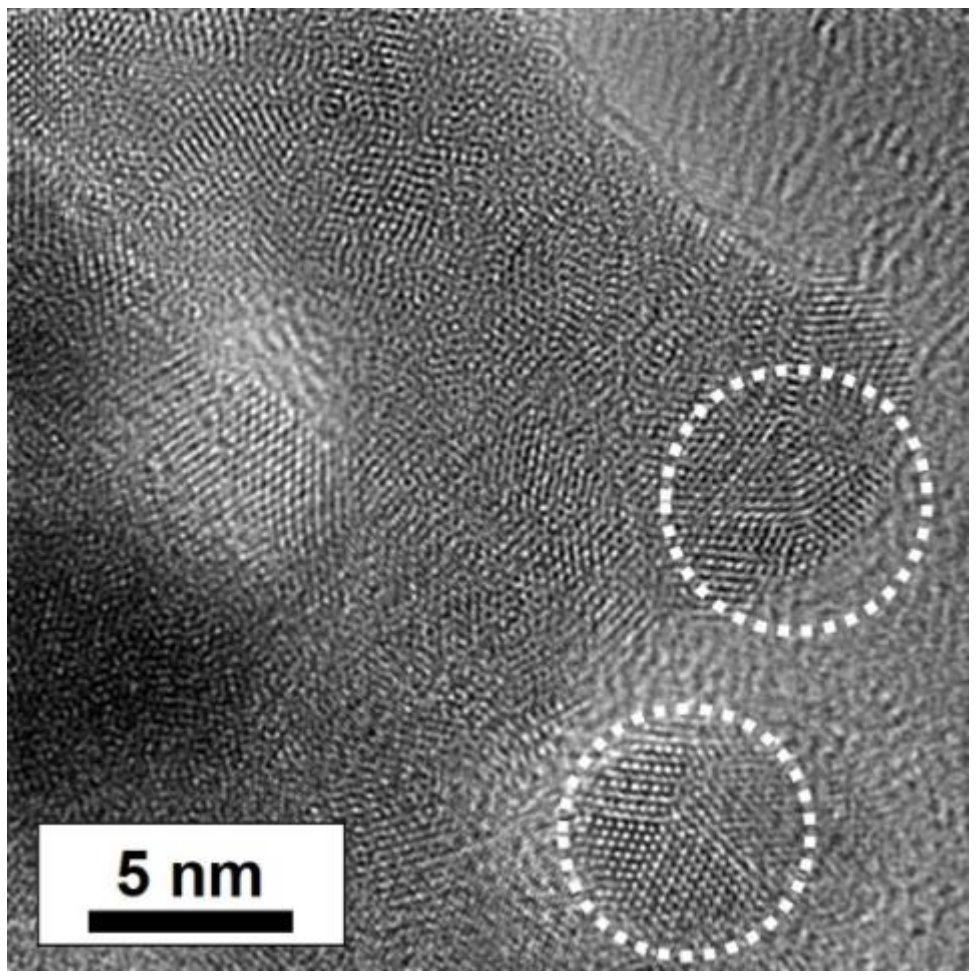

**Figure S18.** HR-TEM image of Pd-$B_{40}$ nano-materials.

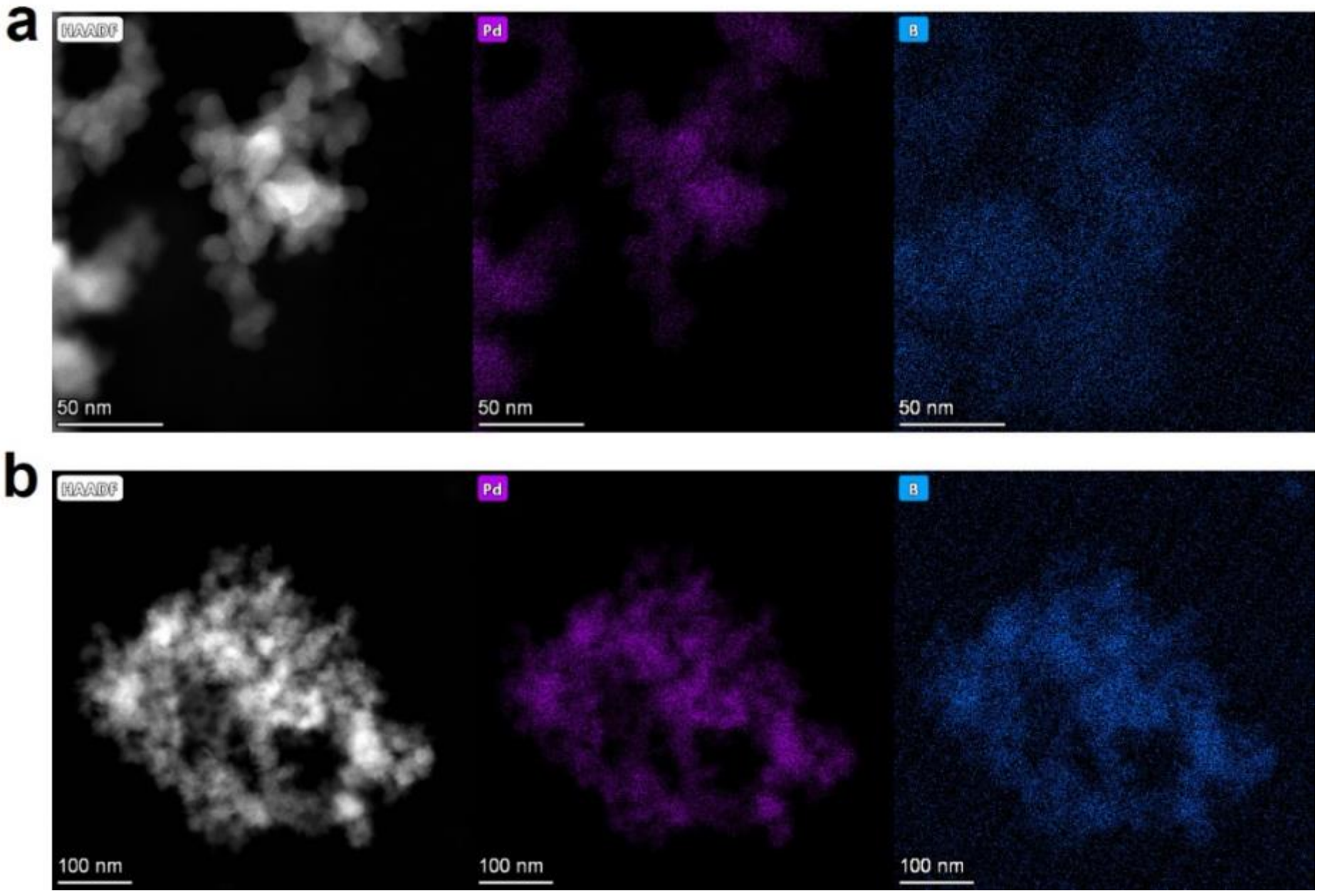

**Figure S19.** HAADF-STEM images and corresponding EDS maps of Pd (purple) and B (blue) for (a) Pd-$B_{40}$ and (b) Pd-$B_{0.1}$.

**B binding sites on Pd.** We perform DFT calculations to investigate the impact of B modifications on the H and OH adsorption behavior on Pd(111) surfaces. First, the stability of B adsorbates on pristine Pd(111) surface is studied (Table S2). The B binding energy for various high-symmetry adsorption sites (top, bridge, FCC and HCP sites on the surface; octahedral and two different tetrahedral sites in the sub-surface) on a Pd(111) surface, is calculated using Equation 1 and plotted as a function of B coverage in Figure 1a. The most stable binding sites are the 3-fold hollow sites (*i.e.*, HCP and FCC sites) for the surface and the octahedral site for the sub-surface. Since the binding energy $E_{\mathrm{b}}$ of the octahedral site is 1.25 eV/(B atom) [1.75 eV/(B atom)] larger than that of the FCC site at 0.25 ML (1 ML), in thermodynamic equilibrium B atoms will strongly prefer to adsorb in the octahedral sub-surface sites. This is in agreement with previous theoretical studies of B adsorption on Pd(111).[28,29]

**The chemical potential shift of B in solution.** To better understand the effect of the experimental conditions for synthesizing MNAs, the shift in the chemical potential of B with respect to the rhombohedral α-phase when the atoms originate from the solution, $\Delta\mu_{\mathrm{ads}}$, is evaluated as[30]

$$\Delta\mu_{\mathrm{B}}\left(c_{\mathrm{BH}_4^-}, T, p\mathrm{H}\right) = \mu_{\mathrm{BH}_4^-} - 4 \cdot \mu_{\mathrm{H}} - \mu_e$$

where $\mu_{\mathrm{BH}_4^-}$, $\mu_{\mathrm{H}}$, and $\mu_e$ are the chemical potentials of $\mathrm{BH}_4^-$ , H, and the electron, respectively. $\mu_{\mathrm{BH}_4^-}$ is evaluated as follows:

$$\mu_{\mathrm{BH}_4^-} = \Delta_f G^\circ(\mathrm{BH}_4^-) - \Delta_f G(\mathrm{BH}_4^-)\left(c_{\mathrm{BH}_4^-}, T\right)$$

where $\Delta_f G^\circ(\mathrm{BH}_4^-)$ and $\Delta_f G(\mathrm{BH}_4^-)$ are the formation energies of $\mathrm{BH}_4^-$ ion in the standard state and at a given temperature and ion concentration, respectively. To determine $\Delta_f G^\circ(\mathrm{BH}_4^-)$ with respect to the rhombohedral α-phase of B, we use the standard reduction potentials for the tabulated reactions involving $BH_4^-$ as given in Ref.[31]:

$$H_2BO_3^- + H_2O + 3e^- \rightleftharpoons B + 4OH^-, \qquad U° = -1.79V$$

$$H_2BO_3^- + 5H_2O + 8e^- \rightleftharpoons BH_4^- + 8OH^-, \qquad U° = -1.24V$$

and follow the procedure described in Ref.[30]. From the two reactions above, $\Delta_f G°(BH_4^-)$ is calculated as:

$$\Delta_f G°(BH_4^-) = -1/8\, eU_2^{\circ} + \Delta_f G°(H_2BO_3^-) - 3\Delta_f G°(H_2O),$$

where $\Delta_f G°(H_2BO_3^-) = 1/3\, eU_1^{\circ} + 3\Delta_f G°(H_2O) + \mu_e^{\mathrm{SHE}} + k_{\mathrm{B}}T \ln c_0$ and $\Delta_f G°(H_2O)$ is the formation energy of water at the standard state which is tabulated as -2.46 eV from Ref.[30]. $\mu_e^{\mathrm{SHE}}$ is the absolute electron chemical potential (i.e. referred to vacuum) of the standard hydrogen electrode (SHE), which is -4.44 eV[32] and $c_0$ is the reference concentration (55.55 mol/l, considering that 1 l of water contains 55.55 mol $H_2O$ molecules). $\Delta_f G(BH_4^-)$ is evaluated based on the experimental concentration of $BH_4^-$ and the temperature as $-k_{\mathrm{B}}T \ln \frac{c_{BH_4^-}}{c_0}$.

Following Ref. [30], $\mu_{\mathrm{H}}$ is a function of the electron chemical potential, $p$H, and temperature:

$$\mu_H = -(\mu_e^{\mathrm{SHE}} - \mu_e) - k_{\mathrm{B}}T \ln 10 \cdot p\mathrm{H}$$

The electron chemical potential, $\mu_e$, can be obtained at the given experimental condition using the Nernst equation:

$$\mu_e = \mu_e^{\mathrm{SHE}} - k_{\mathrm{B}}T \ln \frac{c_{BH_4^-}}{c_0}$$

Combining all the above equations, the chemical potential shift, $\Delta\mu_{\mathrm{B}}\left(c_{BH_4^-}, T, p\mathrm{H}\right)$, can be written as:

$$\Delta\mu_{\mathrm{B}}(c_{\mathrm{BH_4^-}}, T, p\mathrm{H}) = \Delta_f G°(\mathrm{BH_4^-}) - \Delta_f G(\mathrm{BH_4^-})(c_{\mathrm{BH_4^-}}, T) - \mu_e^{\mathrm{SHE}} + 5k_{\mathrm{B}}T \ln\frac{c_{\mathrm{BH_4^-}}}{c_0} + 4k_{\mathrm{B}}T \ln 10 \cdot p\mathrm{H}$$

**Table S2.** Calculated binding energies of B on Pd(111) with respect to the rhombohedral α-phase of B for different B coverages and binding sites

| **$E_b$ [eV/(B atom)]** | | | | |
|---|---|---|---|---|
| **Θ (ML)** | **0.25** | **0.5** | **0.75** | **1** |
| **Top** | **2.16** | **-** | **-** | **2.27** |
| **Bridge** | **-** | **-** | **-** | **1.77** |
| **FCC** | **0.06** | **0.41** | **0.23** | **1.38** |
| **HCP** | **-0.01** | **0.38** | **0.21** | **1.38** |
| **Tetra-down** | **-0.12** | **-** | **-0.11** | **0.19** |
| **Tetra-up** | **-0.54** | **-** | **-0.21** | **0.12** |
| **Octahedral** | **-1.24** | **-1.13** | **-0.78** | **-0.37** |

**Table S3.** The calculated chemical potential shift of B in solution with respect to the rhombohedral α-phase of B at given experimental conditions and tabulated literature data

| **eV/(B atom)** | **Pd-$B_{0.1}$** | **Pd-$B_1$** | **Pd-$B_{40}$** |
|---|---|---|---|
| $\Delta\mu_{\mathrm{B}}$ | **-0.37** | **-0.01** | **0.56** |
| $\Delta_f G°(\mathrm{BH_4^-})$ | **-4.78** | | |
| $\Delta_f G(\mathrm{BH_4^-})$ | **0.28** | **0.22** | **0.13** |

B formation energy and concentration in Pd FCC bulk. To compute the thermodynamic equilibrium concentration of B atoms in Pd nanoparticles, the formation energy of B in the Pd FCC bulk system is calculated based on the equation:

$$E_{\mathrm{f}} = E_{\mathrm{tot}}^{\mathrm{B/Pd\ FCC}} - E_{\mathrm{tot}}^{\mathrm{Pd\ FCC}} - E_{\mathrm{tot}}^{\alpha-\mathrm{B}}$$

where $E_{\mathrm{tot}}^{\mathrm{B/Pd\ FCC}}$, $E_{\mathrm{tot}}^{\mathrm{Pd\ FCC}}$, and $E_{\mathrm{tot}}^{\alpha-\mathrm{B}}$ are the DFT total energies of Pd FCC bulk supercells [i.e., from (1×1×1) to (4×4×4)] with a single interstitial B defect in the octahedral site, the pure Pd FCC bulk and the α-rhombohedral B bulk phase. The calculated formation energies are listed in Table S4. Plotting the formation energies as a function of the corresponding B concentrations (Figure S20), shows a drastic increase in the formation energies, which even become positive (i.e. endothermic reaction), as the B concentrations increases. This is a consequence of the stronger B-B interactions at higher concentrations, which significantly reduces the stability of the system. The thermodynamic equilibrium concentration of B incorporated into a Pd nanoparticle can be estimated by calculating the concentration at the zero formation energy [i.e., $c_B(E_{\mathrm{f}} = 0)$], which is 12.7 B at. % in our calculations. This matches the value of the experimentally observed $Pd_6B$ phase (14.3 B at. %) where B occupies the octahedral site in the Pd FCC host matrix[33]. The close proximity of these two values also demonstrates that experimental protocols, such as using $Pd\text{-}B_{40sc}$, allow us to overcome the kinetic limitations of the wet synthesize approach that prevented to achieve high B concentrations.

**Table S4.** A list of Pd FCC bulk supercells of different sizes, containing B ($N_B$) and Pd ($N_{Pd}$) atoms and a corresponding B concentration [$c_B = N_B / (N_B + N_{Pd})$]. The corresponding calculated formation energy of B in Pd is shown, as well.

| Supercell size | $N_B$ | $N_{Pd}$ | $c_B$ (at. %) | $E_f$ [eV/(B atom)] |
|---|---|---|---|---|
| 1×1×1 | **1** | **4** | **20** | **0.79** |
| 2×2×2 | **1** | **32** | **3.0** | **-1.06** |
| 3×3×3 | **1** | **108** | **0.9** | **-1.47** |
| 4×4×4 | **1** | **256** | **0.4** | **-1.47** |

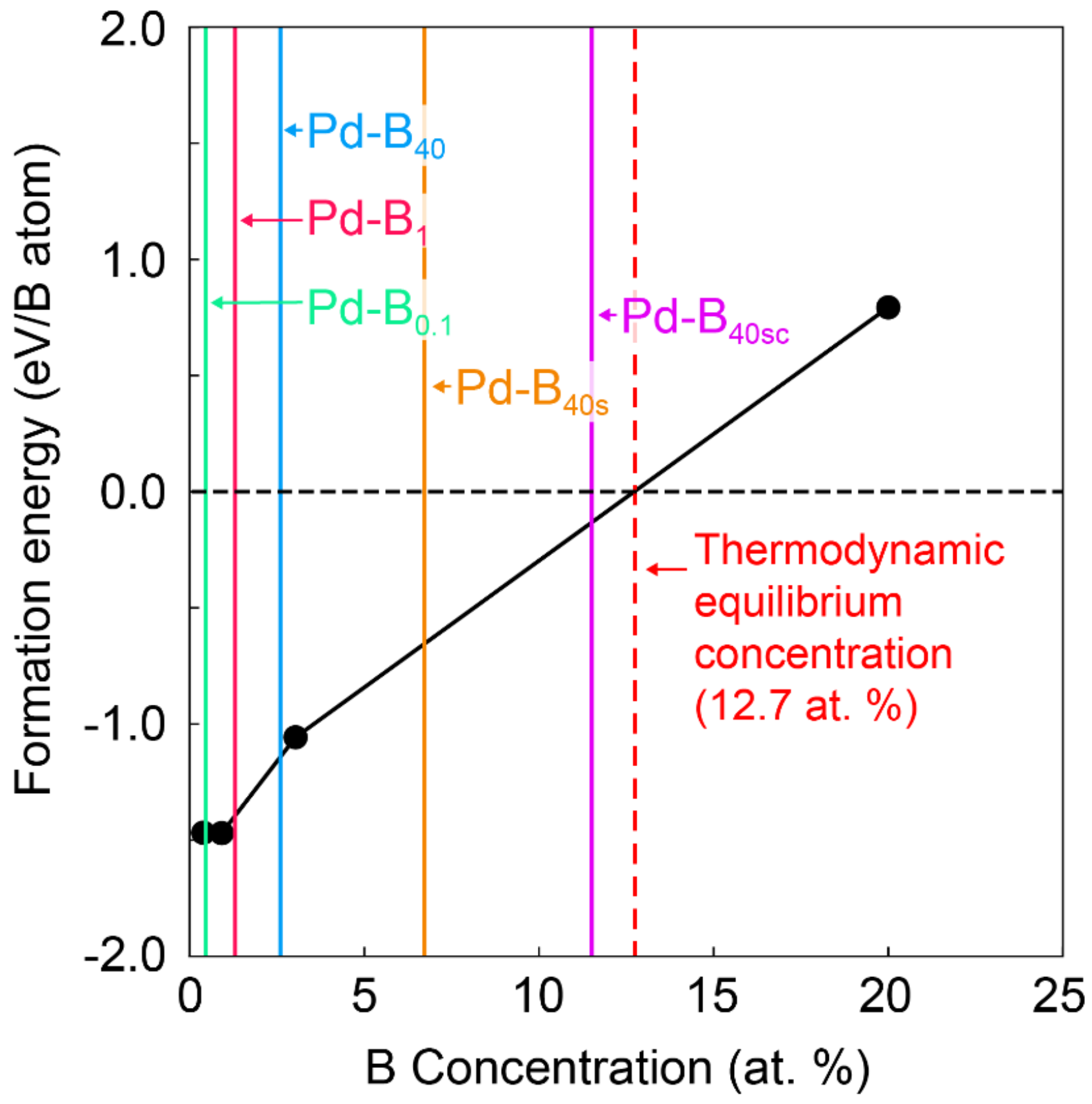

**Figure S20.** The calculated formation energy of a B atom in an Pd FCC bulk as a function of B concentration with respect to the B bulk reference phase (α-rhombohedral phase). The experimentally measured B concentrations in Pd nanoparticles for different conditions (i.e., Pd-$B_{0.1}$, Pd-$B_1$, Pd-$B_{40}$, Pd-$B_{40s}$, and Pd-$B_{40sc}$) are indicated by the corresponding vertical color lines. The approximated thermodynamic equilibrium concentration of B in Pd FCC bulk is indicated by the red dashed vertical line and denoted by the red text in the plot.

## Atomic composition analysis of each Pd-$B_x$ sample.

**Table S5.** Summary of overall atom counts of each element in Pd-$B_x$ nano-materials APT samples.

| **Element (# of collected atoms)** | **Pd-$B_{0.1}$** | **Pd-$B_1$** | **Pd-$B_{40}$** |
|---|---|---|---|
| B | 16749 | 22156 | 58235 |
| C | 17583 | 4996 | 8500 |
| O | 40094 | 20976 | 86598 |
| Na | 207 | 373 | 951 |
| Ni | 13216544 | 310407 | 2288239 |
| Pd | 3600257 | 1661626 | 2170321 |
| **B/(Pd+B) (at.%)** | 0.4631 ±0.0017 | 1.3158 ±0.0081 | 2.6131 ±0.0075 |

*C, O are originated from citric acid used as a buffer acid for electroplating. Also, Na may stem from impurities from chemicals we used (*e.g.* $NaBH_4$). Note that the standard error ($\sigma_s$) is calculated based on a binomial distribution: $\sigma_s = (c_{B}*(1-c_B)/n_t)^{1/2}$, where $n_t$ is the total number of the solute & solvent ions and $c_B$ is the concentration in atomic fraction of a solute (B).

**XRD results.** X-ray diffraction pattern (XRD) is performed for as-synthesized Pd nanomaterials to check their crystallinities. Figure S21 shows XRD patterns of sample Pd-$B_{0.1}$, Pd-$B_1$, and Pd-$B_{40}$. No B or $BO_x$ related peak is found[34] and no boride related peak is found[35]. All samples are in a clear fcc phase of Pd with sharp signals at (111), (200), (220), (311), and (222): the powder diffraction file (PDF) 00-046-1043. However, there is a noticeable trend that Pd-$B_1$ peaks slightly shift to a lower angle and eventually Pd-$B_{40}$ peaks become broader and shifts to a lower angle. The patterns at 40$^o$ show clear indications. Boarder peaks in Pd-$B_{40}$ sample can be interpreted by the presence of complex phases within the nano-aerogel structure. At room temperature, a pure Pd lattice parameter is reported to be 0.3890 nm[36]. From Bragg's law, the lattice parameters, $a_{cal}$, of Pd-$B_{0.1}$, Pd-$B_1$, and Pd-$B_{40}$ are calculated to be 0.3894, 0.3902, and 0.3915 nm, respectively, suggesting that the lattice dilation from the interstitial element, B, could result the peak shifts[37].

M. Beck et al. proposed that chemical interaction between Pd and B plays an important role in the formation of fcc Pd-B bulk and determined a linear function of unit cell parameters of interstitial solid solution of Pd with solute B content as follows[38]:

$$a(x^f_B) = 0.3892 + 0.07697\,(x^f_B) \qquad \text{Equation S1}$$

, where a is a lattice constant of Pd in nm and $x^f_B$ is atomic fraction of B in Pd. The calculated Pd lattice parameters (black) according to Eq. S1 are to be 0.3896, 0.3902, 0.3912 nm for Pd-$B_{0.1}$, Pd-

$B_1$, and Pd-$B_{40}$, respectively, which well match to the Pd lattice parameter values (blue) from the XRD results at B content less than 3 at.%.

In Figure S22, the trendline (blue) shows non-linearity and deviates from the linear function acquired from a bulk Pd-B system. This could be due to nano-aerogel size effects as it also has been seen in copper and titanium dioxide nanoparticle systems[39]. When the Pd-$B_{40s}$ and Pd-$B_{40sc}$ nanoparticles are synthesized with controlled kinetics to dope more B, their average thickness slightly decreases that resulted from their slow growth. Therefore, two contributions counteracting each other could result to overall lattice parameter changes –interstitial atom (B) effect and size effect. Moreover, the interference fringe summation with the XRD peak is not taken account might raise the deviations that is critical to small particles. In overall, the level of Pd lattice expansion increases as B content increases.

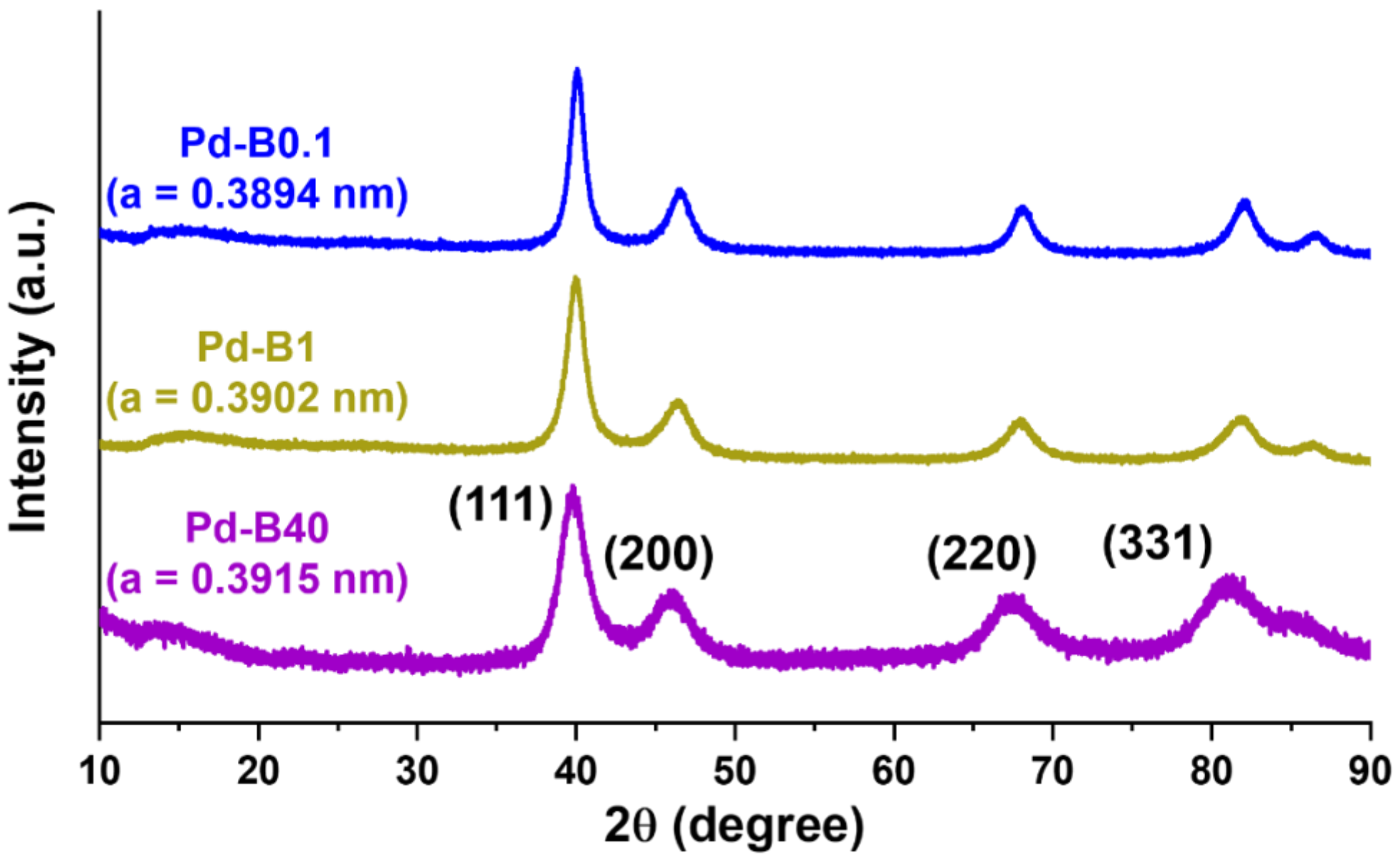

**Figure S21.** XRD patterns of as-synthesized Pd nanomaterials: Pd-$B_{0.1}$ (blue), Pd-$B_1$ (yellow), and Pd-$B_{40}$ (purple).

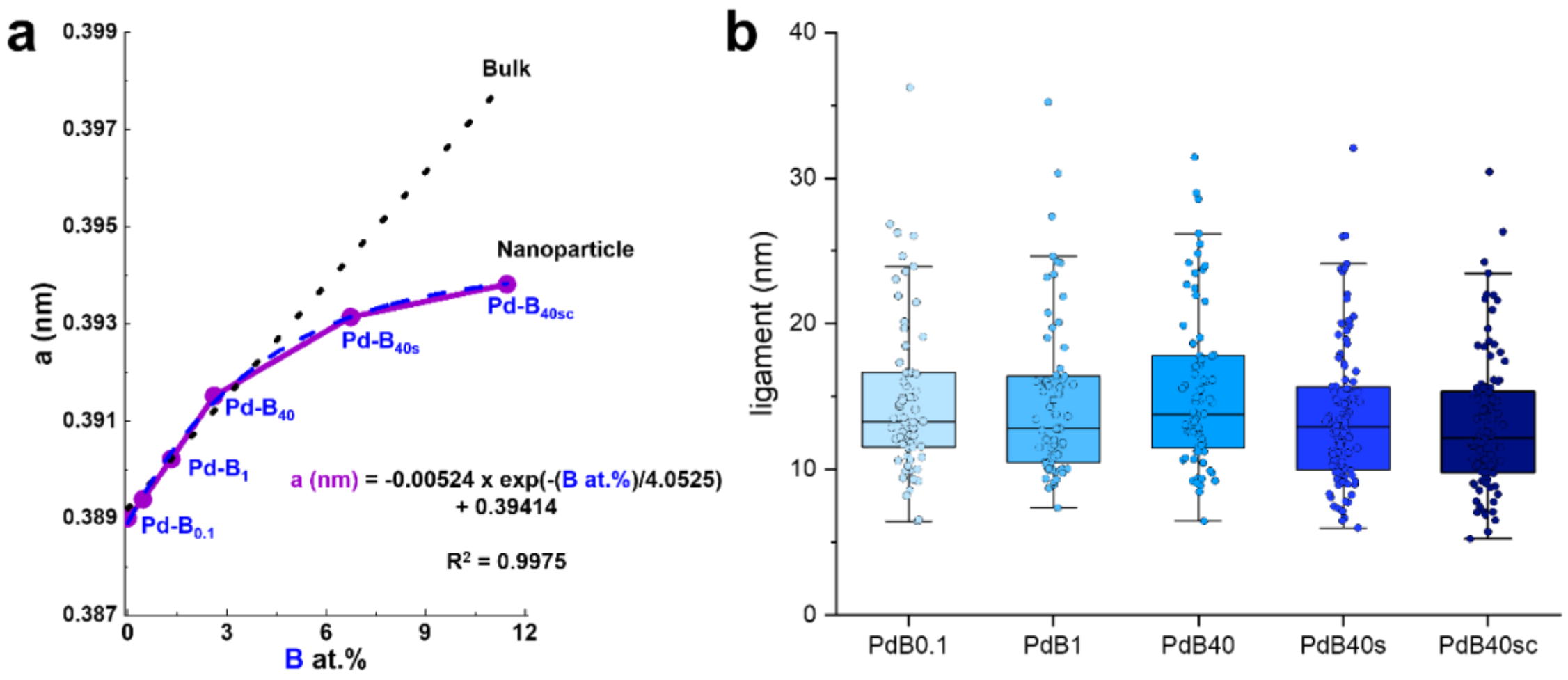

**Figure S22.** (a) A plot of calculated lattice constants versus B composition from the XRD experiment (purple line) and the Beck et al. calibration linear fit (black dotted line). The blue dotted line represents an empirical function of Pd-B nanoparticle system. (b) The average width of >70 ligaments from (HR)- and (S)-TEM images of each sample.

## Characterization of Pd-$B_{40s}$ nano-materials.

As-prepared sodium borohydride solution (same amount as Pd-$B_{40}$ sample) was instilled into a Pd precursor solution (0.01 M). After bubble stopped, the nanoparticles were collected with a centrifuge and rinsed thrice times to remove any impurities residuals. The collected black powder was dried and contained in a vacuum desiccator. Figure S23a and S23b show HR-TEM images of the as-synthesized nanoparticles. No difference in morphology and size is found with other samples. However, in the XRD result, all Pd-$B_{40s}$ FCC-related peaks shifts to low angles indicating a lattice expansion attributed to interstitial B incorporation (see Figure S23c). The calculated lattice constant is 0.3931 nm which is higher value than the lattice constant of Pd-$B_{40s}$ (0.3915 nm).

APT specimen was prepared and measured following the same protocol described above. In Figure 23d, the reconstructed Pd-$B_{40s}$ nanomaterials also exhibit a complex aerogel structure and a thickness of ~15 nm in good correlation with the TEM results. Overall B composition in the reconstructed Pd-$B_{40s}$ nanoparticles is calculated to be 6.7 at.% which is ~3 times more B amount than that of Pd-$B_{40s}$.

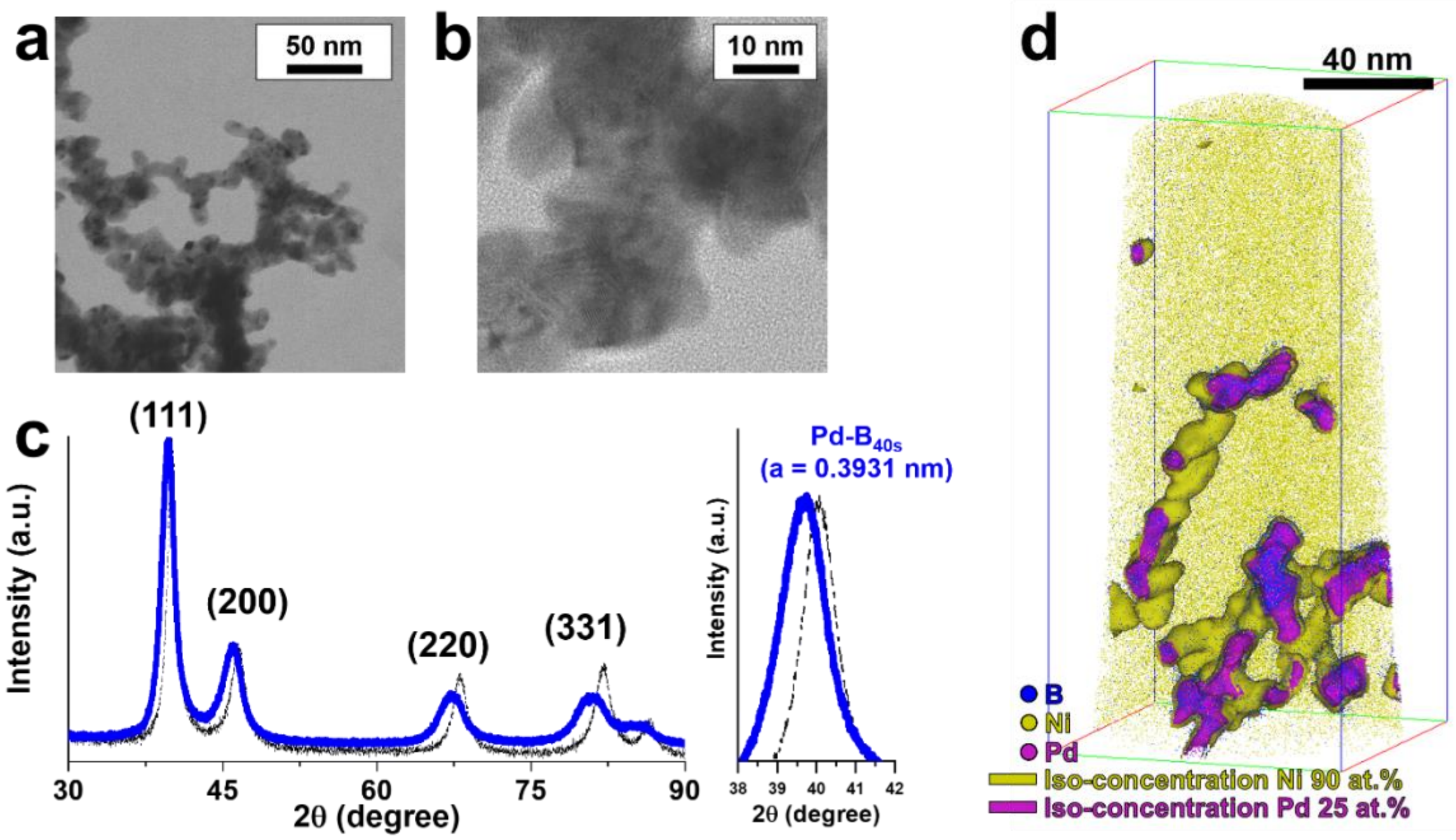

**Figure S23.** (a) and (b) HR-TEM images of Pd-$B_{40s}$ nano-materials. (c) XRD pattern of Pd-$B_{40s}$ nano-materials with Pd-$B_{0.1}$ for comparison (dotted line). Inset shows the detailed information at Pd (111) peak signal. Notice that all peaks of Pd-$B_{40s}$ are shifted to left. (d) 3D atom maps of Pd-$B_{40s}$ nano-materials encapsulated in Ni.

## Characterization of Pd-$B_{40sc}$ nano-materials.

As same condition with the Pd-$B_{40s}$ synthesis, Pd-$B_{40sc}$ nanoparticles was synthesized at cooled environment (~5 $^{o}$C) throughout the reaction. After bubble stopped, the gels were collected using a centrifuge and rinsed with distilled water for surface impurities removals. The acquired dataset from APT is presented in Figure S24d and the summary of atomic compositions of Pd-$B_{40x}$ system is shown in Table S6.

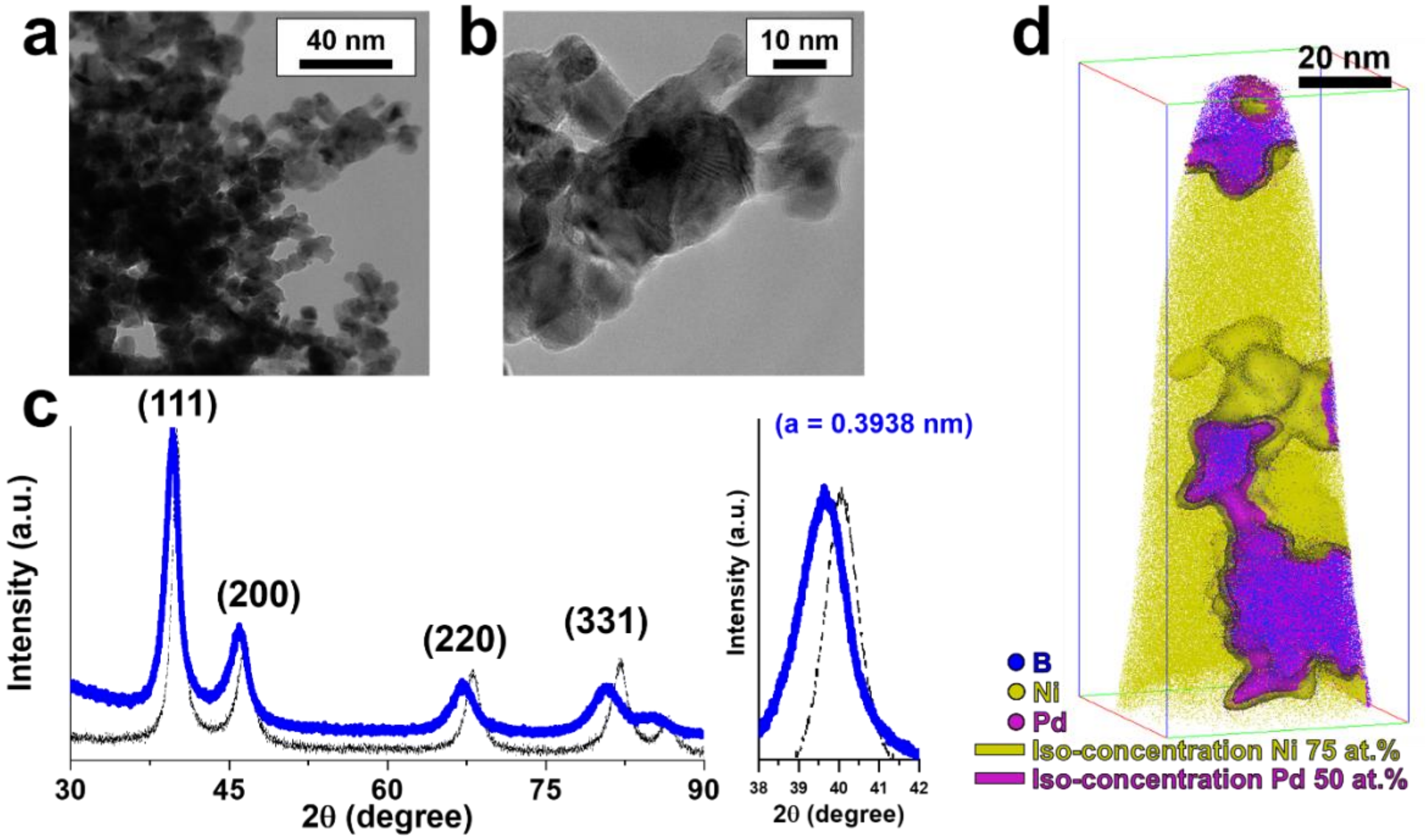

**Figure S24.** (a) and (b) HR-TEM images of Pd-$B_{40sc}$ nano-materials. (c) XRD pattern of Pd-$B_{40sc}$ nano-materials with Pd-$B_{0.1}$ for comparison (dotted line). Inset shows the detailed information at Pd (111) peak signal. Notice that all peaks of Pd-$B_{40sc}$ are shifted to left. (d) 3D atom maps of Pd-$B_{40sc}$ nano-materials encapsulated in Ni.

**Table S6.** Summary of overall atom counts of each element in Pd-$B_{40x}$ from APT.

| Element (# of collected atoms) | Pd-$B_{40}$ | Pd-$B_{40s}$ | Pd-$B_{40sc}$ |
|---|---|---|---|
| B | 58235 | 58766 | 226637 |
| C | 8500 | 11032 | 1501 |
| O | 86598 | 53491 | 16647 |
| Na | 951 | 1134 | 175 |
| Ni | 2288239 | 56341563 | 7181144 |
| Pd | 2170321 | 813610 | 1750825 |
| **B/(Pd+B) (at.%)** | 2.6131 ±0.0075 | 6.7363 ±0.0268 | 11.4610 ±0.2265 |

Electronic structures of Pd-B. To assess the filling of the antibonding states, we determine the position of the *d*-band centre calculated by Equation 2 (in the main text) relative to the Fermi energy. Numerous studies confirm that this is the simplest possible measure for the position of the *d*-states and a good indicator of bonding strength[40], since the antibonding states of the adsorbate are always located above the metal *d*-states. Consequently, a higher energy of the *d*-band centre results in the stronger chemical bond.  Hence, we investigate coverage dependent changes in the *d*-band centre for B located in the octahedral sub-surface site based on the *d*-band model.

There are two contributions to these changes – a strain effect[41] and an alloying effect[42]. These effects often counteract each other: A lattice expansion, as induced by placing B in the sub-surface region of the Pd(111) surface and seen in Figure S25a, leads to a narrowing of the *d*-band and an upward shift of the *d*-band centre; the electronic interactions due to "alloying", as achieved by the incorporation of B in the sub-surface region of Pd, induce on the other hand a broadening of the *d*-band and a downward shift of the *d*-band centre.  Indeed, Figure 25b reveals a large depletion of electrons from the top Pd surface layer, due to the absorption of B, indicating a strong interaction between the B *p*-states and the Pd *s*- and *d*-states. In order to identify which effect dominates, we look (Figure 25c) at the coverage dependent change in the density-of-states (DOS) of the $B_{octa}$/Pd(111) system, as the B coverage is varied. This figure reveals a clearly seen broadening of the DOS as the B coverage is increased, due to the extension of the valence band tail down to approximately -6 eV, which in turn results in a monotonous down shift of the *d*-band centre. This analysis reveals two important points. Firstly, the alloying effect of B in Pd overrides the strain effect, which is in good agreement with previous studies[29,43]. Secondly, the downward shift of the *d*-band centre induced by the presence of B sub-surface atoms should weaken the binding energy of adsorbates on the B containing Pd(111) surface.

**Table S7.** Calculated binding energies of a system with H adsorbed on-surface and B adsorbed in an octahedral subsurface site ($B_{octa}$) on Pd(111) for different combinations of on-surface/subsurface coverages and H adsorption sites. Coverages Θ for H and B are varied from 0 to 1 ML, with 1 ML corresponding to a coverage in which the number of adsorbate atoms of one kind (either H or B) equals the number of Pd atoms in the surface layer.

| $E_b$ (eV/(H atom)) for | | | | |
|---|---|---|---|---|
| **Adsorption site H** | **Θ ML ($H_{on-surface}$ / $B_{octa}$)** | | | |
| | **0.25 H/ 0 $B_{octa}$** | **0.5 H/0 $B_{octa}$** | **0.75 H/0 $B_{octa}$** | **1 H/0 $B_{octa}$** |
| **Top** | **-0.02** | **0.01** | **0.04** | **0.08** |
| **FCC** | **-0.53** | **-0.52** | **-0.51** | **-0.52** |
| **HCP** | **-0.50** | **-0.48** | **-0.48** | **-0.48** |
| **Adsorption site H** | **Θ ML ($H_{on-surface}$ / $B_{octa}$)** | | | |
| | **0.25 H / 0.25 $B_{octa}$** | **0.5 H/0.25 $B_{octa}$** | **0.75 H/0.25 $B_{octa}$** | **1 H/0.25 $B_{octa}$** |
| **Top** | **0.00** | **-** | **-** | **0.17** |
| **FCC** | **-0.49** | **-0.49** | **-0.46** | **-0.37** |
| **HCP** | **-0.22** | **-0.34** | **-0.43** | **-0.36** |
| **Adsorption site H** | **Θ ML ($H_{on-surface}$ / $B_{octa}$)** | | | |
| | **0.25 H/1 $B_{octa}$** | **0.5 H/1 $B_{octa}$** | **0.75 H/1 $B_{octa}$** | **1 H/1 $B_{octa}$** |
| **Top** | **0.06** | **-** | **0.51** | **0.81** |
| **FCC** | **-0.16** | **-** | **0.09** | **0.32** |
| **HCP** | **-0.14** | **-0.14** | **-0.01** | **0.10** |

**Table S8.** Values reported in the literature for the binding energy ($E_b$) of H adsorbed on an on-surface adsorption site on the Pd(111) surface in units of eV/H atom.

| **xc-Functional** | **rPBE**[44] | **PW91**[44] | **GGA**[45,a] | | **GGA**[46] | |
|---|---|---|---|---|---|---|
| **H coverage $\Theta_H$** | **0.11 ML** | **0.11 ML** | **1 ML** | **0.33 ML** | **1 ML** | **0.25 ML** |
| **FCC** | **-0.34** | **-0.64** | **-0.35** | **-0.53** | **-0.39** | **-0.44** |
| **HCP** | | | | | **-0.34** | **-0.41** |
| **xc-Functional** | **LDA**[47] | | **PW91**[48] | | **GGA**[49,b] | **Exp.**[50] |
| **H coverage $\Theta_H$** | **1 ML** | **0.25 ML** | **1 ML** | **0.25ML** | **0.67 ML** | **< 0.7 ML** |
| **FCC** | **-0.43** | **-0.69** | **-0.5** | **-0.55** | **-0.50** | **-0.45±0.02** |
| **HCP** | **-0.37** | **-0.65** | | | **-0.45** | |

a. Used GGA: the exchange energy from Becke[51] and the correlation energy from Perdew[52]
b. Used GGA: the exchange-correlation functional based on the quantum Monte Carlo simulations by Ceperley and Alder[53] as parameterized by Perdew and Zunger[54] to include GGA correction[55]

**Table S9.** Calculated *d*-band centers of Clean Pd(111) without any adsorbate, a H-free Pd(111) with B adsorbed in an octahedral subsurface sites ($B_{octa}$), and a system with H adsorbed on-surface and B adsorbed in ($B_{octa}$) on Pd(111) for different combinations of on-surface/subsurface coverages and H adsorption sites. Coverages Θ for H and B are varied from 0 to 1 ML

| ***d*-band center (eV) for** | | | | |
|---|---|---|---|---|
| **Clean Pd(111)** | –1.52 | | | |
| **H-free Pd(111)** | **Θ ML ($B_{octa}$)** | | | |
| | **0.25** | **0.50** | **0.75** | **1.00** |
| **Octa** | **-1.55** | **-1.61** | **-1.79** | **-1.90** |
| **Adsorption site H** | **Θ ML ($H_{on-surface}$ / $B_{octa}$)** | | | |
| | **0.25 H / 0 $B_{octa}$** | **0.5 H/0 $B_{octa}$** | **0.75 H/0 $B_{octa}$** | **1 H/0 $B_{octa}$** |
| **Top** | **–1.54** | **–1.53** | **–1.57** | **–1.55** |
| **FCC** | **–1.53** | **–1.51** | **–1.52** | **–1.48** |
| **HCP** | **–1.53** | **–1.52** | **–1.52** | **–1.49** |
| **Adsorption site H** | **Θ ML ($H_{on-surface}$ / $B_{octa}$)** | | | |
| | **0.25 H / 0.25 $B_{octa}$** | **0.5 H/0.25 $B_{octa}$** | **0.75 H/0.25 $B_{octa}$** | **1 H/0.25 $B_{octa}$** |
| **Top** | **-1.57** | **-** | **-** | **-1.63** |
| **FCC** | **-1.57** | **-1.55** | **-1.60** | **-1.66** |
| **HCP** | **-1.55** | **-1.54** | **-1.58** | **-1.64** |
| **Adsorption site H** | **Θ ML ($H_{on-surface}$ / $B_{octa}$)** | | | |
| | **0.25 H/1 $B_{octa}$** | **0.5 H/1 $B_{octa}$** | **0.75 H/1 $B_{octa}$** | **1 H/1 $B_{octa}$** |
| **Top** | **-1.90** | **-** | **-1.97** | **-2.02** |
| **FCC** | **-1.89** | **-** | **-1.83** | **-2.04** |
| **HCP** | **-1.87** | **-1.93** | **-1.93** | **-2.01** |

**Table S10.** Calculated binding energies of a system with OH adsorbed on-surface and B adsorbed in an octahedral subsurface site ($B_{octa}$) on Pd(111) for different combinations of on-surface/subsurface coverages and H adsorption sites. Coverages Θ for OH and B are varied from 0 to 1 ML, with 1 ML corresponding to a coverage in which the number of adsorbate atoms of one kind (either OH or B) equals the number of Pd atoms in the surface layer.

| $E_b$ (eV/(OH molecule)) for | | | | |
|---|---|---|---|---|
| **Adsorption site OH** | **Θ ML ($OH_{on-surface}$ / $B_{octa}$)** | | | |
| | **0.25 OH/ 0 $B_{octa}$** | **0.5 OH/0 $B_{octa}$** | **0.75 OH/0 $B_{octa}$** | **1 OH/0 $B_{octa}$** |
| **Top** | **–1.67** | **–2.17** | **–1.88** | **–1.18** |
| **FCC** | **–2.61** | **–** | **–1.94** | **–1.61** |
| **HCP** | **–2.46** | **–2.17** | **–1.86** | **–1.51** |
| **Adsorption site OH** | **Θ ML ($OH_{on-surface}$ / $B_{octa}$)** | | | |
| | **0.25 OH / 0.25 $B_{octa}$** | **0.5 OH/0.25 $B_{octa}$** | **0.75 OH/0.25 $B_{octa}$** | **1 OH/0.25 $B_{octa}$** |
| **Top** | **–** | **–** | **–2.30** | **–1.53** |
| **FCC** | **–** | **–** | **–2.38** | **–1.67** |
| **HCP** | **–2.69** | **–2.22** | **–2.36** | **–1.61** |
| **Adsorption site OH** | **Θ ML ($OH_{on-surface}$ / $B_{octa}$)** | | | |
| | **0.25 OH/1 $B_{octa}$** | **0.5 OH/1 $B_{octa}$** | **0.75 OH/1 $B_{octa}$** | **1 OH/1 $B_{octa}$** |
| **Top** | **–2.55** | **–2.28** | **–** | **–1.37** |
| **FCC** | **–** | **–** | **–2.37** | **–1.58** |
| **HCP** | **–** | **–2.30** | **–2.47** | **–1.67** |
| **Bridge** | **–3.00** | **–** | **–** | **–** |

**Table S11.** Calculated *d*-band centers of systems with OH adsorbed on-surface and B adsorbed in ($B_{octa}$) on Pd(111) for different combinations of on-surface/subsurface coverages and H adsorption sites. Coverages Θ for OH and B are varied from 0 to 1 ML.

| ***d*-band center (eV) for** | | | | |
|---|---|---|---|---|
| **Adsorption site OH** | **Θ ML ($OH_{on\text{-}surface}$ / $B_{octa}$)** | | | |
| | **0.25 OH / 0 $B_{octa}$** | **0.5 OH/0 $B_{octa}$** | **0.75 OH/0 $B_{octa}$** | **1 OH/0 $B_{octa}$** |
| **Top** | –1.39 | –1.44 | –1.47 | –1.45 |
| **FCC** | –1.43 | – | –1.50 | –1.48 |
| **HCP** | –1.42 | –1.46 | –1.48 | –1.44 |
| **Adsorption site OH** | **Θ ML ($OH_{on\text{-}surface}$ / $B_{octa}$)** | | | |
| | **0.25 OH / 0.25 $B_{octa}$** | **0.5 OH/0.25 $B_{octa}$** | **0.75 OH/0.25 $B_{octa}$** | **1 OH/0.25 $B_{octa}$** |
| **Top** | – | – | –1.49 | –1.57 |
| **FCC** | – | – | –1.50 | –1.52 |
| **HCP** | –1.45 | –1.48 | –1.50 | –1.53 |
| **Adsorption site OH** | **Θ ML ($OH_{on\text{-}surface}$ / $B_{octa}$)** | | | |
| | **0.25 OH/1 $B_{octa}$** | **0.5 OH/1 $B_{octa}$** | **0.75 OH/1 $B_{octa}$** | **1 OH/1 $B_{octa}$** |
| **Top** | –1.74 | –1.68 | – | –1.72 |
| **FCC** | – | – | –1.74 | –1.77 |
| **HCP** | – | –1.77 | –1.74 | –1.77 |
| **Bridge** | –1.72 | – | – | – |

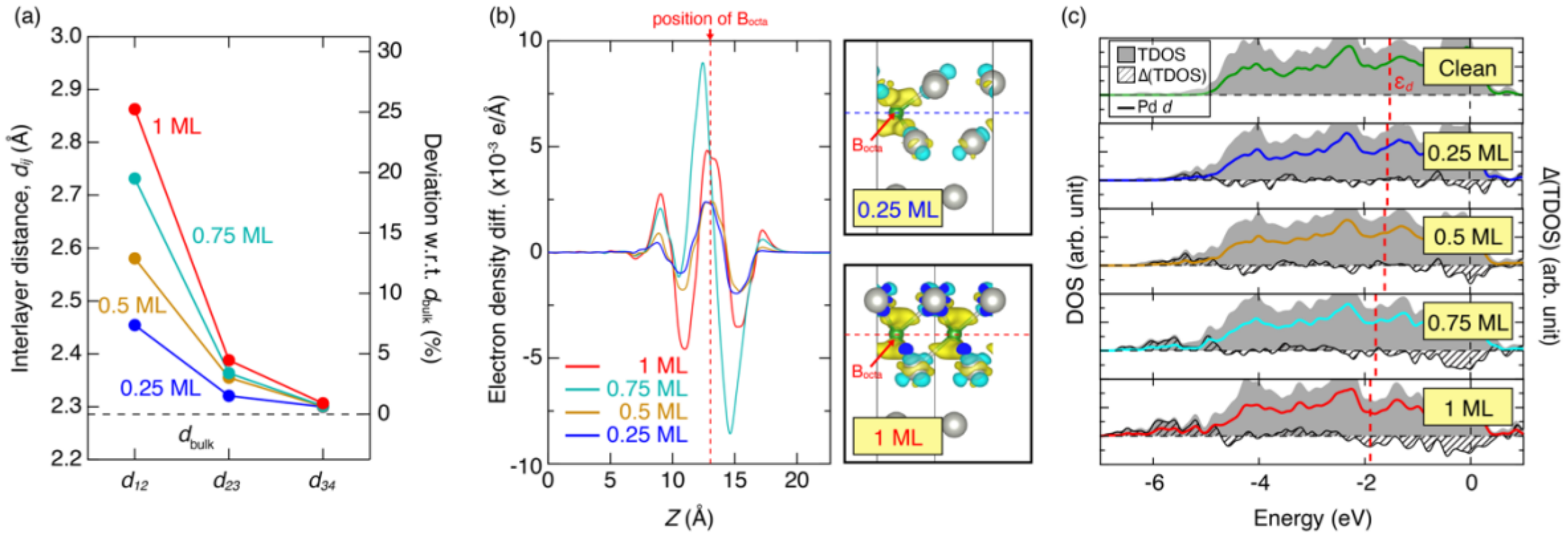

**Figure S25.** (a) Interlayer distance between the outmost layers *i* and *j* layers, $d_{ij}$ of a Pd(111) slab. The black horizontal dashed line indicates the bulk interlayer spacing. Blue, orange, cyan and red solid lines depict geometries in which the coverage of B, adsorbed in the octahedral sub-surface site between the two outermost Pd layers [$B_{Octa}$/Pd(111)], is increasing from 0.25 ML to 1 ML in ¼ ML steps. (b) The Planar-averaged electron density difference (left panel) between the overall system $B_{octa}$/Pd(111) and its constituents Pd(111) and B (*i.e.*, $\Delta\rho = \rho(B_{octa}/Pd) - \rho(Pd) - \rho(B)$ ) shown perpendicular to the surface as a function of the *z*-coordinate is shown to the left. The vertical red dashed line represents the position of the B atom in the octahedral site. The right panel shows 3D plots of the electron density difference for 0.25 ML (right-top panel) and 1 ML (right-bottom panel). Grey and green balls indicate Pd and B atoms, respectively. Yellow and cyan regions indicate electron accumulation and electron depletion regions, respectively. (c) Density-of-states (DOS) shown for increasing B coverage in the octahedral subsurface site ($B_{octa}$) on Pd(111) starting from a B-free Pd(111) surface (top panel) to a Pd(111) surface with 1 ML $B_{Octa}$ (bottom panel). The grey and the black hashed regions show the total DOS and the difference between this total DOS and the total DOS of the clean surface, respectively. Colored solid lines show the contribution of Pd *d*-states. The Fermi energy is set to 0 eV. Vertical red lines show the d-band centre values calculated using Equation 2.

# Preliminary Studies

Synthesizing Pd nanoparticle seeds. We synthesized the Pd nanoparticle seed following the N. R. Jana et al. method[56]. 0.005 M of Pd ions was prepared in 10 mL of distilled water mixed with 0.005 M of citric acid. Then 0.5 mL of $NaBH_4$ solution (0.1 M) was dropped immediately in the as-prepared Pd ions complex solution, which is the same mole ratio to $Pd-B_1$ nano-aerogel synthesis. After bubbles stopped, as-synthesized Pd seeds were collected using a centrifuge (12000 rpm, 15 min) followed by re-dispersing in distilled water. This process was repeated three times for removal of residuals following the modified method[57].

After the synthesis, Pd seeds (< 2 nm in size) and dendritic Pd nanoparticles (~20 nm) are observed with high-angle annular dark-field scanning transmission electron microscopy, HADDF-STEM images (see Figure S26a and S26b). The number density of the seeds is much greater than that of the nano-dendrite. During the growth, some seeds could be aggregated to form dendrite structure[58] as a citric acid (surfactant) is known as a weak binding ligand[59].

To investigate presences of B in Pd nanoparticle seeds, chemistry analysis using electron spectroscopy techniques are also performed. Overall energy-dispersive X-ray spectroscopy (EDS) spectrum of the Pd seeds shows a noticeable peak at 0.188 keV which corresponds to B (see Figure S26c). Furthermore, from high resolution HAADF-STEM image and corresponding EELS mapping for B Kα results, we detected strong B signals on an edge of dendritic Pd nanoparticles (see Figure S27). Unfortunately, these do not provide enough information of location (surface or sub-surface) and atomic concentration of B in nanoparticle system. Therefore, in order to precisely map B in 3D and to measure its concentration in atomic-level, atom probe tomography can be used to investigate B incorporation in the seeds.

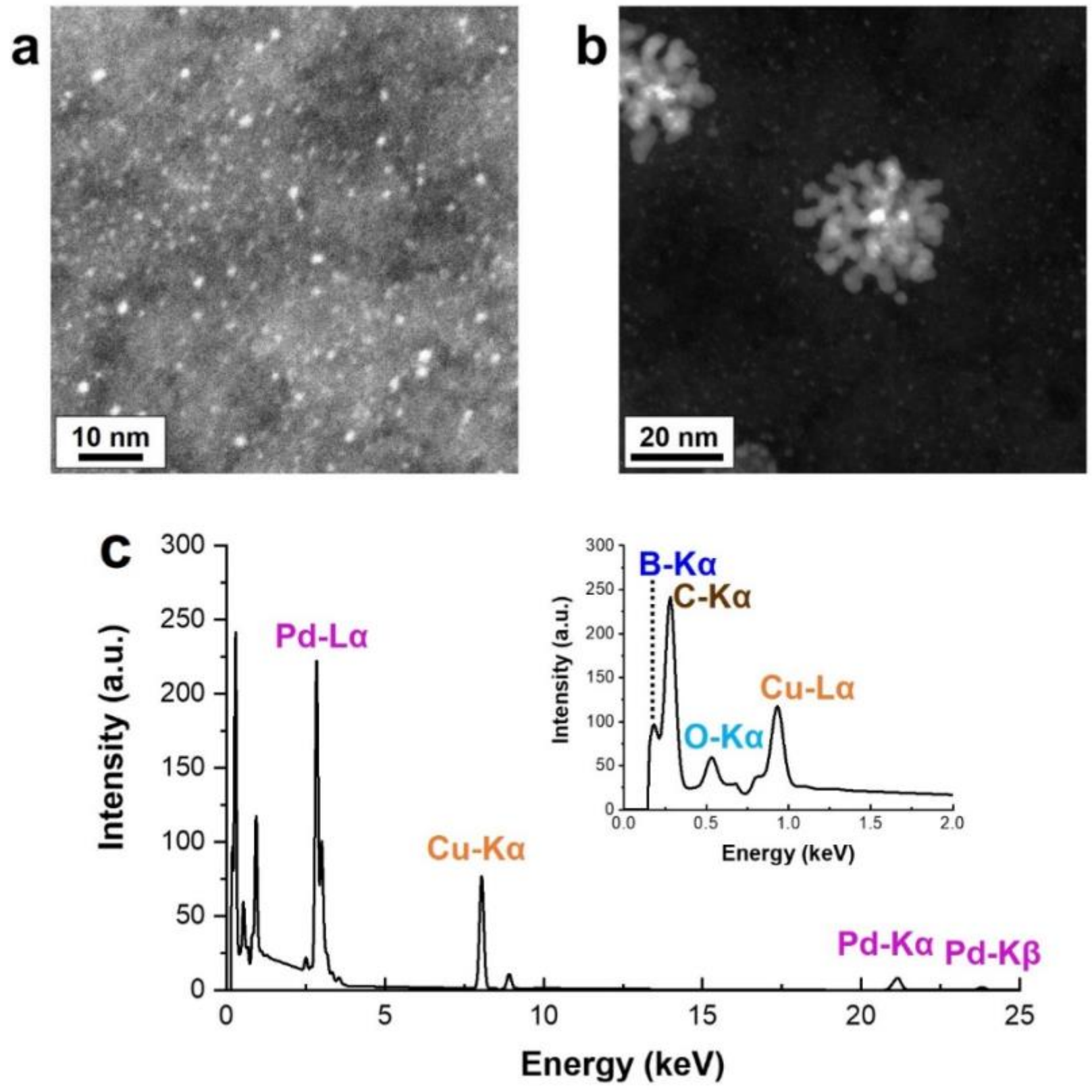

**Figure S26.** (a,b) HAADF-STEM images of as-synthesized Pd nanoparticle seeds and nano-dendrites. (c) EDS spectra of as-synthesized Pd seeds. Inset image shows the EDS spectrum range between 0 to 2 keV.

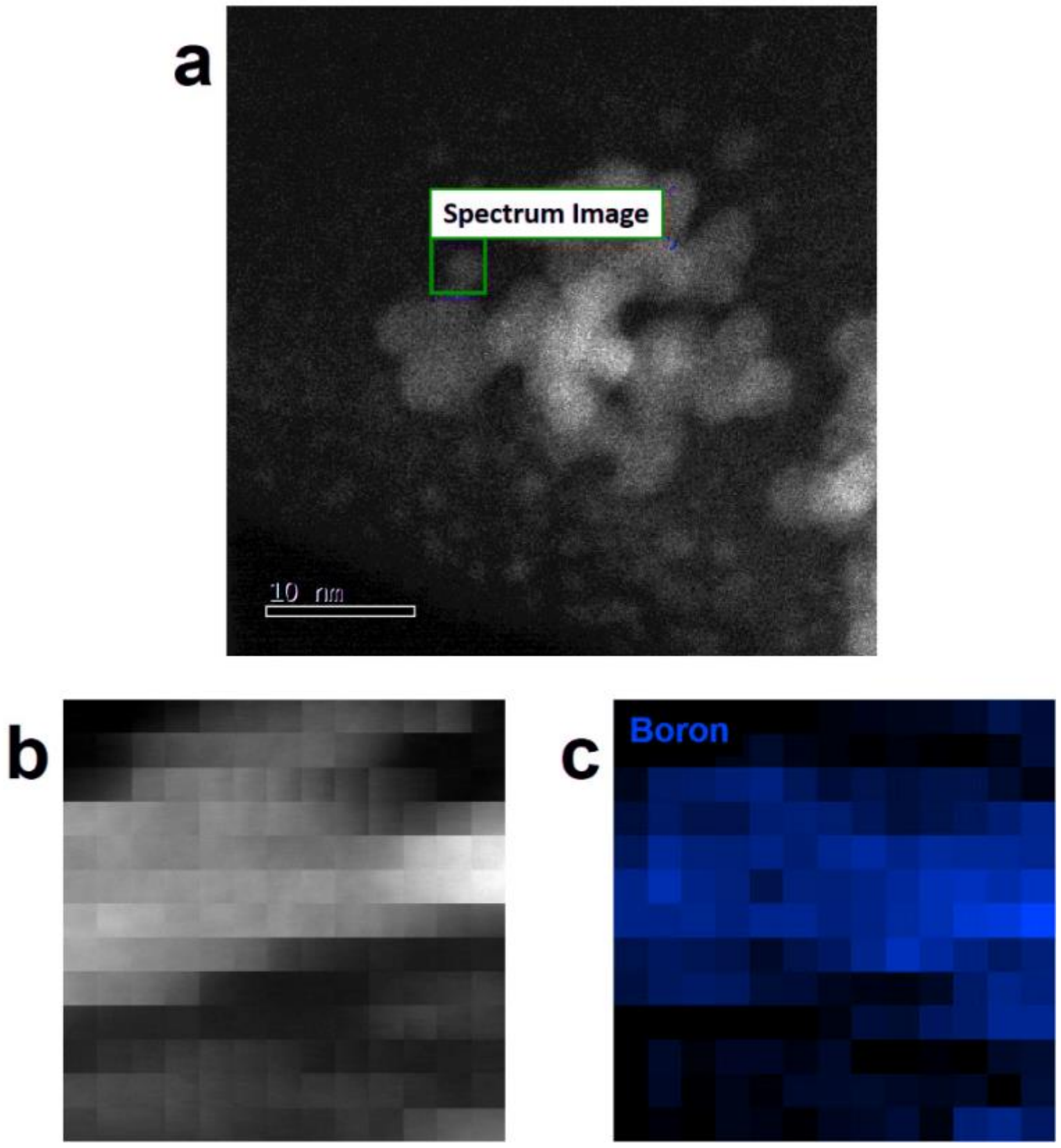

**Figure S27.** (a) HAADF-STEM, (b) zoomed in ADF-STEM image, and (c) corresponding STEM-EELS B maps of nano-dendrites.